\documentclass[11pt, letterpaper, fleqn]{article}
\usepackage{amsmath, amssymb, amsthm, mathtools, tikz, todonotes, bbm, hyperref, fullpage, cleveref}
\usepackage{complexity}
\usepackage{keytheorems}

\newcommand{\numberthis}{\addtocounter{equation}{1}\tag{\theequation}}

\newkeytheorem{theorem}[parent=section]
\newkeytheorem{corollary}[sibling=theorem]
\newkeytheorem{lemma}[sibling=theorem]

\newkeytheorem{proposition}[sibling=theorem]

\theoremstyle{definition}
\newtheorem{definition}[theorem]{Definition}

\newcommand{\eqnref}[1]{(\ref{#1})}
\newcommand{\bits}{\{-1, 1\}}
\newlang{\AND}{AND}
\newcommand{\sr}{\mathrm{rank}_{\pm}}
\newlang{\OR}{OR}
\newclass{\UPP}{UPP}
\newcommand{\UPPcc}{\UPP^\cc}
\newcommand{\PPcc}{\PP^\cc}

\newlang{\PM}{PM}
\newlang{\MAJ}{MAJ}

\newlang{\HS}{HS}
\newlang{\CHS}{CHS}

\renewcommand{\E}{\mathop{\mathbb{E}}}
\renewcommand{\R}{\mathbb{R}}
\newcommand{\Z}{\mathbb{Z}}
\newcommand{\N}{\mathbb{N}}

\newcommand{\sgn}{\operatorname{sgn}}
\newcommand{\dd}{\mathrm{d}}

\newcommand{\bra}[1]{\{#1\}}
\newcommand{\pmone}{\bra{-1, 1}}

\renewcommand{\deg}{\mathrm{deg}}

\newcommand{\sdeg}{\mathrm{deg}_{\pm}}

\newcommand{\mathify}[1]{\ifmmode{#1}\else\mbox{$#1$}\fi}

\newcommand{\rank}{\mathrm{rank}}

\usepackage[dvipsnames]{xcolor}
\hypersetup{
    colorlinks,
    linkcolor=BrickRed,
    citecolor=NavyBlue,
    urlcolor=Cerulean,
    linktoc=page
}

\usepackage{cite}
\title{UPP$^{\text{cc}}$ is Not Closed Under Intersection}

\date{October 27, 2026}

\begin{document}

\author{%
Mark Bun \\
Boston University \\
\texttt{mbun@bu.edu}
\and
Mandar Juvekar \\
Boston University \\
\texttt{mandarj@bu.edu}
}
\maketitle

\begin{abstract}
The sign rank of a ${\pm 1}$-valued matrix $A$ is the minimum rank of a real matrix that agrees with $A$ entrywise in sign. For every sufficiently large $d$, we give an explicit matrix $A$ with $2^{\Theta(d^2)}$ rows and columns and sign rank $\Theta(d)$ such that $A \land A$ has sign rank $2^{\Theta(d)}$. Here, $A \land A$ is the communication matrix of the problem obtained by evaluating $A$ on disjoint inputs and taking the conjunction of the results. As the logarithm of sign rank characterizes unbounded-error communication complexity, this implies that the communication class $\UPPcc$ is not closed under intersection.

The starting point for our lower bound is Forster's method, which relates the sign rank of a matrix to the average margin of representations of its rows and columns by points and halfspaces. We study a continuous analog for conjunctions of inner-product sign tests on the unit sphere. Rotational symmetry connects this problem to polynomial sign representations, yielding an exponentially small upper bound on the average margin. This bound holds even when sign constraints are relaxed near the sign tests' decision boundary, allowing us to transfer it to explicit finite sign matrices arising from intersections of integer halfspaces.

\end{abstract}

\newpage
\tableofcontents
\newpage

\section{Introduction}\label{s: intro}

The unbounded error model of communication complexity~\cite{PS86} is both powerful and mathematically fascinating. Let $F : \mathcal{A} \times \mathcal{B} \to \{-1, 1\}$ be a known Boolean function defined over finite sets $\mathcal{A}, \mathcal{B}$.
Here, we identify with $-1$ and 1 the Boolean values ``true'' and ``false'' respectively.
Alice holds an input $a \in \mathcal{A}$, Bob holds an input $b \in \mathcal{B}$, and both parties have access to an unlimited sequence of private random bits. Their goal is to compute $F(a, b)$ with any probability strictly greater than $1/2$ (i.e., strictly better than random guessing) by exchanging as few bits of communication as possible. Let $U(F)$ denote the minimum number of bits exchanged in any such communication protocol, and denote by $\UPPcc$ the class of communication problems $F$ for which $U(F)$ is polylogarithmic in the input size.

The class $\UPPcc$ is a communication complexity analog of the Turing machine class $\PP$, which captures languages decidable in polynomial time by randomized algorithms that are correct with probability strictly greater than $1/2$. The flip side to these classes' permissive success conditions is that standard error reduction techniques fail, making it challenging to prove closure properties. It was a major open question whether $\PP$ is closed under intersection, until a breakthrough of Beigel, Reingold, and Spielman~\cite{BRS95} demonstrated this to be the case. (Their proof also shows that a different communication complexity analog, $\PPcc$, is also closed under intersection.) However, the situation for $\UPPcc$ has remained unresolved. To our knowledge, Halstenberg and Reischuk~\cite{HalstenbergR88} were the first to explicitly ask whether $\UPPcc$ is closed under intersection with several other authors reiterating the question more recently~\cite{GPW18, Sherstov2018HardestHalfspaceTalk, BunMT21, bun-tha:approximate-degree}.\footnote{See also~\cite{HatamiHPTZ22, GoosHIS25} who asked whether the UPP communication of an intersection can be bounded \emph{at all} in terms of that of the original problem.}

Our main result gives a negative answer to this question. For a function $F : \mathcal{A} \times \mathcal{B} \to \{-1, 1\}$, define $F \land F: \mathcal{A}^2 \times \mathcal{B}^2 \to \bits$ by $(F \land F)((a_1, a_2), (b_1, b_2)) = F(a_1, b_1) \land F(a_2, b_2)$, where the symbol $\land$ denotes Boolean AND. We show that there is an infinite family of explicit functions $F$ on $n$-bit inputs such that $U(F) = O(\log n)$, but $U(F \land F) = \Omega(\sqrt{n})$. (These parameters are obtained by taking $M = 2^{cd}$, and hence $n = \Theta(d^2)$ in the following statement.)

\begin{definition}
        For natural numbers $M, d$, let $\mathcal{A} = \mathcal{B} = \{-M, \dots, -1, 0, 1, \dots, M\}^d$ and define the \emph{bounded integer halfspace problem} $\HS_{M, d} : \mathcal{A} \times \mathcal{B} \to \{-1, 1\}$ by $\HS_{M, d}(a, b) = \sgn(\langle a, b\rangle - 1/2)$.%
        \footnote{That is, $\HS_{M, d}(a, b) = -1$ if $\langle a, b \rangle \le 0$ and $\HS_{M, d}(a, b) = 1$ if $\langle a, b \rangle > 0$. The affine shift of $-1/2$ only serves to disambiguate its behavior when $\langle a, b \rangle = 0$.}
\end{definition}

\begin{theorem}[Restatement of~\Cref{thm:HS-lower-bound}] \label{thm:halfspace-main}
For all $M, d$ we have $U(\HS_{M, d}) = O(\log d)$. Meanwhile, there is a constant $c$ such that if  $M \ge 2^{cd}$, we have  $U(\HS_{M, d} \land \HS_{M, d}) = \Omega(d)$. Note that the input length of the problem is $n = d \cdot \lceil \log (2M + 1)\rceil$.
\end{theorem}

Prior to this work, the only examples showing that $U$ can increase under intersection at all were given by Bun, Mande, and Thaler~\cite{BunMT21}, who described a family of functions $F$ such that $U(F) = O(\log n)$, but $U(F \land F) = \Theta(\log^2 n)$.

Both our unbounded error communication upper bound for $\HS_{M, d}$ and our lower bound for $\HS_{M, d} \land \HS_{M, d}$ are tight for $M = 2^{\Theta(d)}$, so our separation cannot be improved for this family of functions. (See Section~\ref{sec:tightness} for details.) We also show how to obtain the same separation for the same halfspace evaluation problem, but where $\mathcal{A}$ and $\mathcal{B}$ consist of $2^{\Theta(d^2)}$ random points on the unit sphere in $d$ dimensions. These problems arise as alternative discretizations of the same continuous problem that we describe below.

The rest of this introduction consists of a technical overview of the proof of~\Cref{thm:halfspace-main} (Section~\ref{sec:tech-overview}), a description of selected applications (Section~\ref{sec:intro-apps}), a discussion of related work (Section~\ref{sec:related}), and open problems (Section~\ref{sec:open}).

\subsection{Technical Overview} \label{sec:tech-overview}

\subsubsection{Sign rank} In their paper introducing the unbounded error communication model, Paturi and Simon~\cite{PS86} identified an elegant matrix-analytic characterization of $U(F)$. For a Boolean matrix $A \in \{-1, 1\}^{\mathcal{A} \times \mathcal{B}}$, define the sign rank of $A$, denoted $\sr(A)$, to be the minimum rank of a real matrix $R \in \mathbb{R}^{\mathcal{A} \times \mathcal{B}}$ such that $A_{a, b} R_{a, b} > 0$ for every $(a, b) \in \mathcal{A} \times \mathcal{B}$. That is, the sign rank of a Boolean matrix is the least rank of a real matrix that agrees with it entrywise in sign. Letting $A$ denote the communication matrix of a two-party function $F$, i.e., $A_{a, b} = F(a, b)$, Paturi and Simon showed that
\[U(F) = \log (\sr(A)) \pm O(1).\]
The sign rank of a matrix $A$ also has a natural geometric interpretation: We say $A$ is realized by an $r$-dimensional \emph{linear arrangement} if there exist collections of unit vectors $\{u_a\}_{a \in \mathcal{A}} \subseteq \mathbb{S}^{r-1}$ and $\{v_b\}_{b \in \mathcal{B}} \subseteq \mathbb{S}^{r-1}$ such that $\sgn \langle u_a, v_b \rangle = A_{a, b} $ for every pair $a, b$. That is, if we view $\{u_a\}$ as a collection of points and $\{v_b\}$ as a collection of vectors on the unit sphere $\mathbb{S}^{r-1} \subseteq \mathbb{R}^r$, we have that $A_{a, b} = 1$ iff $u_a$ lies on the ``positive'' side of the hyperplane $v_b$ is normal to. The sign rank of $A$ is the minimum dimension $r$ in which  such a linear arrangement exists.

It is straightforward from its definition that the communication matrix of $\HS_{M, d}$ has sign rank at most $d+1$, and hence it has unbounded error communication complexity $O(\log d)$. The rest of this overview is devoted to showing that the sign rank of this matrix increases exponentially under two-fold intersection.

In a breakthrough result establishing an exponential lower bound on the sign rank of Hadamard matrices, Forster~\cite{Forster02} discovered how to prove lower bounds on the dimension of linear arrangements via upper bounds on their \emph{average margin}. For a distribution $\mu$ over $\mathcal{B}$, the average margin of arrangement $\{u_a\}, \{v_b\}$ with respect to $\mu$ is
\[\min_{a \in \mathcal{A}} \mathbb{E}_{b \sim \mu}[|\langle u_a, v_b\rangle|].\]
So an arrangement has small average margin if there exists a row $a$ such that the average distance from $u_a$ to the hyperplane defined by $v_b$ is small under the distribution $\mu$.
Following a generalization of Forster's argument due to~\cite{HatamiHPTZ22}, we extend this definition to the average margin of a sign matrix as
\[
m^{\mathrm{avg}}(A) = \inf_{\mu} \sup_{\{u_a\}, \{v_b\}} \min_{a \in \mathcal{A}} \E_{b \sim \mu}[|\langle u_a, v_b \rangle|].
\]
Here, the outer infimum is taken over distributions $\mu$ on $\mathcal{B}$ and the supremum is taken over linear arrangements of \emph{arbitrary} dimension that realize $A$.
Forster's argument and the refinement of~\cite{HatamiHPTZ22} show that low-dimensional arrangements imply arrangements with large average margin; quantitatively,
\[\sr(A) \ge \frac{1}{m^{\mathrm{avg}}(A)}\]
for every Boolean matrix $A$.

Forster's original paper bounded the average margin based on the spectral norm of $A$, and subsequent work building on his technique developed further matrix-analytic refinements~\cite{forster2002smallest, forsteretal, linial2007complexity, sherstovsymm}.

\subsubsection{A continuous formulation} \label{sec:overview-continuous}

In this work, we take a more geometric view to upper bound the average margin of the communication  matrix of $\HS_{M, d} \land \HS_{M, d}$. The first key step is to formulate a continuous version of the problem, which we denote by $\CHS_d \land \CHS_d$. Each input to a copy of $\CHS_d$ is an arbitrary point on the unit sphere $\mathbb{S}^{d-1}$, rather than an integer vector. Instead of being a pair of discrete vector assignments, a linear arrangement for $\CHS_d \land \CHS_d$ is now a pair of functions $f, g : (\mathbb{S}^{d-1})^2 \to \mathbb{S}^{r-1}$ such that 
\begin{equation} \label{eq:continuous-arrangement}
    \sgn \langle f(x_1, x_2), g(y_1, y_2) \rangle = \sgn \langle x_1, y_1 \rangle \land \sgn \langle x_2, y_2\rangle
\end{equation}
for all $(x_1, x_2), (y_1, y_2) \in (\mathbb{S}^{d-1})^2$.
The quantity $\mathbb{E}[|\langle f(X_1, X_2), g(Y_1, Y_2) \rangle|]$ is an upper bound on the average margin of this arrangement, with respect to the uniform distribution, where the expectation is taken over $X_1, X_2, Y_1, Y_2 \sim \mathbb{S}^{d-1}$ independently and uniformly.
The upshot of this formulation is that it lets us exploit rotational symmetry of the sphere. Specifically, the right-hand side of~\eqnref{eq:continuous-arrangement} depends only on the inner products $s_1 := \langle x_1, y_1\rangle$ and $s_2 := \langle x_2, y_2\rangle$. Define a function $H : [-1, 1]^2 \to [-1, 1]$ by taking $H(s_1, s_2)$ to be the expected value of $\langle f(X_1, X_2), g(Y_1, Y_2)\rangle$ over random spherical inputs with the prescribed inner products $s_1 = \langle X_1, Y_1\rangle$ and $s_2 = \langle X_2, Y_2\rangle$. The definition of this bivariate ``symmetrized'' function immediately shows that it has the following useful properties:
\begin{description}
\item[(a)] $\sgn H(s_1, s_2) = \sgn s_1 \land \sgn s_2$ for all $s_1, s_2 \in [-1, 1]$ by \eqref{eq:continuous-arrangement}, and
\item[(b)] $\mathbb{E}[|H(S_1, S_2)|] = \mathbb{E}[|\langle f(X_1, X_2), g(Y_1, Y_2) \rangle|]$ and is hence an upper bound on the average margin of the arrangement $f, g$ with respect to the uniform distribution.
The former expectation is taken over $S_1, S_2$ independently distributed as the inner product of two uniform points on the sphere. Our sign rank lower bound of $2^{\Omega(d)}$ on the original discrete problem will eventually follow from showing that this quantity is always at most $2^{-\Omega(d)}$.
\end{description}
A final key property of $H$, which we prove in Theorem~\ref{thm:sym-derivatives} and Corollary~\ref{cor:sym-taylor}, is that the averaging process smooths it out. In particular, $H$ is continuously differentiable and has bounded derivatives through order $O(d)$. Hence, by Taylor's theorem, it is approximable by its low-degree Taylor polynomials:
\begin{description}
    \item[(c)] The degree-$(d/2 - 2)$ Taylor polynomial $T$ of $H$ satisfies
    \[|H(s_1, s_2) - T(s_1, s_2)| \le 2^{-\Omega(d)}\]
    for all $s_1, s_2 \in [-1/16, 1/16]$.
\end{description}

Property (c), however, comes into immediate tension with property (a). This is because polynomials do a poor job of representing the ``lower-left quadrant'' indicator $Q_2(s_1, s_2) := \sgn s_1 \land \sgn s_2$. In particular, we claim that it is impossible for a polynomial $p(s_1, s_2)$, no matter its degree, to sign-represent $Q_2(s_1, s_2)$, i.e., $\sgn p(s_1, s_2) = Q_2(s_1, s_2)$ for all nonzero $s_1, s_2 \in [-1, 1]$.
To see this, suppose such a sign-representing polynomial $p$ existed. Let $\ell$ be the lowest (total) degree of any term appearing in $p$, and let $p_\ell(s_1, s_2)$ be its degree-$\ell$ homogeneous part.
In the neighborhood of $0$, the polynomial $p$ is dominated by $p_\ell$, so $p_\ell$ itself must sign-represent $Q_2$ wherever it is nonzero. But the homogeneity of $p_\ell$ makes this impossible:
\begin{itemize}
    \item If $\ell$ is even, then $p_\ell(s_1, s_2) = p_\ell(-s_1, -s_2)$, whereas the upper-right and lower-left quadrants of $[-1, 1]^2$ require opposite signs.
    \item If $\ell$ is odd, then $p_\ell(s_1, s_2) = -p_\ell(-s_1, -s_2)$, whereas the upper-left and lower-right quadrants both require a positive sign.
\end{itemize}

It is tempting to conclude from this discussion that we have just reached a contradiction, and that the function $H$, and therefore the original arrangement $f, g$ we started with, simply cannot exist.
The way out, however, is actually for the degree-$(d/2-2)$ Taylor polynomial for $H$ to be uniformly small---in fact, the stronger version of property (c) expressed in~\Cref{cor:sym-taylor} implies it must be identically zero.
The approximation guarantee then entails that $H$ itself is small near the origin, i.e., $|H(s_1, s_2)| \le 2^{-\Omega(d)}$ for all $s_1, s_2 \in [-1/16, 1/16]$. This is now enough to imply an upper bound on the average margin estimate from property (b): Concentration arguments (e.g., \Cref{lem:score-conc-bounds}) tell us that the inner product of uniformly random points on the sphere falls into any constant-size interval around the origin with probability $1 - 2^{-\Omega(d)}$. Hence, $\mathbb{E}[|H(S_1, S_2)|] = 2^{-\Omega(d)}$, proving the target average margin bound for $\CHS_d \land \CHS_d$.

\subsubsection{Transfer to the discrete problem}

To transfer our continuous average margin bound to one for the original discrete problem, we would like to be able to argue as follows. Start with an arbitrary linear arrangement $\{u_a\}, \{v_b\}$ for the discrete problem $\HS_{M,d} \land \HS_{M,d}$. Extend this to a linear arrangement $f, g$ for  $\CHS_d \land \CHS_d$ by setting $f(x_1, x_2) = u_a$ where $a = (a_1, a_2)$ is the pair of integer vectors closest to $(x_1, x_2)$ upon normalization, and similarly for $g$. The discussion in Section~\ref{sec:overview-continuous} shows that $f, g$ have small average margin, which implies that $\{u_a\}, \{v_b\}$ do as well.

The problem with this argument is that the rounding procedure might not produce a valid arrangement $f, g$ for the continuous problem. Suppose $(x_1, x_2), (y_1, y_2) \in (\mathbb{S}^{d-1})^2$ are points that are extremely close to the ``decision boundary'' where either $\langle x_1, y_1\rangle \approx 0$ or $\langle x_2, y_2\rangle \approx 0$. Then rounding these points might result in discrete points $a, b$ whose inner products have different signs from what we started with. That is, while we are assuming that $\langle u_a, v_b\rangle$ agrees in sign with $\sgn\langle a_1, b_1\rangle \land \sgn\langle a_2, b_2\rangle$, this might not in turn agree with $\sgn\langle x_1, y_1\rangle \land \sgn\langle x_2, y_2\rangle$.

Our solution is to strengthen our average margin bound for the continuous problem $\CHS_d \land \CHS_d$ so that it holds even if the arrangements $f, g$ are allowed to give incorrect signs close to the decision boundary. More precisely, our main technical result is the following.

\begin{theorem}[Special case of~\Cref{thm:relaxation-lower-bound}]
    Let $f, g : (\mathbb{S}^{d-1})^2 \to \mathbb{S}^{r-1}$ be measurable functions such that
    \[\sgn \langle f(x_1, x_2), g(y_1, y_2) \rangle = \sgn \langle x_1, y_1 \rangle \land \sgn \langle x_2, y_2\rangle\]
    whenever $|\langle x_1, y_1\rangle|, |\langle x_2, y_2\rangle| > 2^{-Cd}$ for some universal constant $C > 0$.
    Then 
    \[\mathbb{E}\big[ |\langle f(X_1, X_2), g(Y_1, Y_2) \rangle| \big] \le 2^{-\Omega(d)}\]
    where the expectation is taken over $X_1, X_2, Y_1, Y_2 \sim \mathbb{S}^{d-1}$ independently and uniformly.
\end{theorem}

To establish this upgraded average margin bound, we still have property (c) of the symmetrized function $H$ described in Section~\ref{sec:overview-continuous} at our disposal, as well as an approximate version of property (b) and the weakening of property (a) that holds for all $|s_1|, |s_2| > 2^{-Cd}$. 
The breakdown in our argument arises in our use of the fact that polynomials cannot sign-represent the lower-left quadrant indicator $Q_2$; while that was enough to bound the magnitude of a smooth function $H$ that sign-represents $Q_2$ on the whole square $([-1, 1])^2$, it does not suffice when we only know $H$ sign-represents $Q_2$ on the restricted domain $([-1, -2^{-Cd}] \cup [2^{-Cd}, 1])^2$.

To fix this, we appeal to results from approximation theory~\cite{newman1964approximation, Sherstov13} showing that low-degree polynomials fail to sign-represent $Q_2$ even when each input coordinate is promised to be bounded away from 0. Specifically,~\Cref{lem:strict-sign-lb} shows that no bivariate polynomial of degree $d$ sign-represents $Q_2$ on $([-1, -2^{-Cd}] \cup [2^{-Cd}, 1])^2$ for sufficiently large constant $C$. 
This tells us that the Taylor polynomial $T$ must make some sign errors, and hence both $T$ and $H$ must be small in magnitude at the location of these errors. But this is not enough to guarantee that $H$ is small in an entire constant-size neighborhood of the origin. To derive this stronger conclusion, we further strengthen the approximation theoretic result to control polynomials that may make bounded sign errors when representing $Q_2$. Specifically, we use polynomial interpolation techniques to strengthen~\Cref{lem:strict-sign-lb} as follows.

\begin{lemma}[Consequence of~\Cref{lem:poly-lower-bound}] \label{lem:poly-lb-informal}
    Let $D \geq 1$ be an integer and
    let $p$ be a degree-$D$ bivariate polynomial such that
    \[
        Q_2(s_1, s_2) p(s_1, s_2) \geq -1
    \]
    for all $s_1, s_2 \in [-1, -2^{-C'D}] \cup [2^{-C'D}, 1]$ where $C'$ is a universal constant.
    Then every coefficient of $p$ has magnitude at most $2^{O(D^2)}$.
\end{lemma}

Choosing $D = O(d)$ appropriately, properties~(a) and (c) imply that a rescaling of the degree-$D$ Taylor polynomial for $H$ satisfies the condition of this lemma, and thus has bounded coefficients.
Undoing the scaling then shows that the Taylor polynomial of \emph{lower degree $D/2$} has magnitude $2^{-\Omega(D^2)} \ll 2^{-\Omega(D)}$ on a constant-sized neighborhood of the origin.
Applying the error estimate from Taylor's theorem, this time to the degree-$(D/2)$ Taylor polynomial, yields a modified version of condition~(c) that is sufficient to recover our conclusion that $|H(s_1, s_2)| \leq 2^{-\Omega(D)}$ for all $s_1, s_2 \in [-1/16, 1/16]$ and hence our average margin bound.

\subsection{Applications} \label{sec:intro-apps}

Known connections between $\UPPcc$, sign rank, and other Boolean complexity measures immediately yield several consequences of our results.

\paragraph{Threshold-of-Majority Circuit Size.} In circuit complexity, a threshold gate $t$ on inputs $x_1, \dots, x_n$ computes a function $t(x_1, \dots, x_n) = \sgn(a_1 x_1 + \dots + a_n x_n - \theta)$ for some fixed integer coefficients $a_1, \dots, a_n, \theta$.
It was shown in \cite{forsteretal} that if a Boolean function $F : \{0, 1\}^n \times \{0, 1\}^n \to \{-1, 1\}$ is computed by a depth-$2$ circuit consisting of an arbitrary top threshold gate applied to $s$ bottom threshold gates, each having integer weights $a_i$ of magnitude at most $W$, then $\sr(F) \le O(nsW)$. Since the Majority function is computed by a threshold gate with all weights equal to $1$, this in particular means Threshold-of-Majority circuits computing $F$ require size $s$ at least $\Omega(\sr(F)/n)$. Our main result implies that Threshold-of-Majority circuit size can increase exponentially under intersection.

\begin{corollary}[store=circuit-sep] \label{cor:circuit-sep}
    There is an explicit family of functions $F : \{0, 1\}^{2n} \to \{-1, 1\}$ such that $F$ can be computed by a Threshold-of-Majority circuit of polynomial size, but every Threshold-of-Majority circuit computing $F \land F$ requires size $2^{\Omega(\sqrt{n})}$.
\end{corollary}

\paragraph{Barriers to Learning.} Let $\mathcal{C}$ be a concept class of Boolean functions from a finite domain $\mathcal{X}$ to $\{-1, 1\}$. One can associate a sign matrix $A$ to this class by indexing its rows by functions in $\mathcal{C}$ and its columns by points in $\mathcal{X}$, and by setting each entry to be the row function evaluated on the column point. In the learning theory literature, the sign rank of the matrix $A$ is known as the \emph{dimension complexity} of $\mathcal{C}$, and may be thought of as the smallest dimension $r$ in which the task of learning $\mathcal{C}$ may be reduced to the (well-understood) task of learning halfspaces in $r$ dimensions. More explicitly $\mathcal{C}$ has dimension complexity $r$ if and only if there are $r$ ``feature maps'' $\varphi_1, \dots, \varphi_r : \mathcal{X} \to \mathbb{R}$ such that every function $f \in \mathcal{C}$ may be written as $f(x) = \sgn(a_1 \varphi_1(x) + \dots + a_r \varphi_r(x))$ for some real numbers $a_1, \dots, a_r$.
Upper bounds on dimension complexity underlie state-of-the-art PAC learning algorithms for fundamental concept classes such as DNF, read-once formulas, and intersections of majorities~\cite{KSdnf, Lee09, ambainis2010any, KOS04}. Recent work~\cite{ChornomazMW25} showed that dimension complexity also characterizes the existence of ``exact'' reductions to the considerably broader setting of stochastic convex optimization. Our results show that  intersections of even two halfspaces have exponential dimension complexity, thus ruling out low-dimensional exact reductions in these frameworks.

\begin{corollary}[store=halfspace-learning] \label{cor:halfspace-learning}
    Let $\mathcal{C}_d$ be the concept class consisting of intersections of two halfspaces in dimension $d$. This class has dimension complexity $2^{\Omega(d)}$ even if the points and halfspace weights in $\mathcal{C}_d$ come from an integer grid with each coordinate having bit complexity $O(d)$.

    On the Boolean domain $\{0, 1\}^n$, the class consisting of intersections of two Boolean halfspaces has dimension complexity $2^{\Omega(\sqrt{n})}$.
\end{corollary}

Distribution-free PAC learning intersections of halfspaces is a central and notoriously difficult problem in computational learning theory. It was a longstanding open problem to design an algorithm running in time $2^{o(d)}$ for learning an intersection of two halfspaces, even over the Boolean domain.
A recent breakthrough of Alman, Patel, and Servedio~\cite{AlmanPS26} showed that it is possible to learn an intersection of two real halfspaces in time $2^{\widetilde{O}(\sqrt{d})}$. Their algorithm does not go by way of a dimension complexity upper bound, which our results show is inherent.

\paragraph{Separation Between Sign Rank and Approximate Sign Rank.} In their study of reductions in learning theory, Chornomaz, Moran, and Waknine~\cite{ChornomazMW25} introduced a relaxation of sign rank to capture linear representations that tolerate a small amount of classification error. For a parameter $\varepsilon > 0$, the $\varepsilon$-approximate sign rank of Boolean matrix $A \in \{-1, 1\}^{\mathcal{A} \times \mathcal{B}}$ is the least dimension $r$ in which there is a collection of vectors $\{v_b\}_{b \in \mathcal{B}} \subseteq \mathbb{R}^{r}$ such that for every $a \in \mathcal{A}$ and distribution $\mu$ over $\mathcal{B}$, there exists a vector $u_a \in \mathbb{R}^{r}$ such that $\Pr_{b \sim \mu}[\sgn(\langle u_a, v_b\rangle) \ne A_{a, b}] \le \varepsilon$. Chornomaz, Moran, and Waknine~\cite[Theorem 4]{ChornomazMW25} observed that intersections of majorities yield a quasipolynomial separation between sign rank and $(1/3)$-approximate sign rank, via the lower bound of~\cite{BunMT21}. The question of determining how large the gap can be between sign rank and approximate sign rank was reiterated by~\cite{BinduaHKR26}, who showed that Hadamard matrices also exhibit a quasipolynomial separation. Our work gives a sequence of matrices for which this separation is exponential.

\begin{corollary}[store=approx-vs-exact] \label{cor:approx-vs-exact}
    There is an explicit family of Boolean matrices $A_d$ with $(1/3)$-approximate sign rank at most $2d+1$, but $\sr(A_d) \ge 2^{\Omega(d)}$.
\end{corollary}

\subsection{Related Work} \label{sec:related}

The special case $\HS_{M,3}$ of our bounded integer halfspace problem was studied by \cite{HosseiniHL22}, who used its sign rank upper bound to obtain an essentially optimal separation between sign rank and discrepancy.

In spite of intensive study over the last four decades, our toolkit for proving lower bounds on $\UPPcc$ communication and sign rank is still limited enough to summarize here. VC dimension~\cite{PS86} and (the logarithm of the reciprocal of) the maximum monochromatic rectangle density~\cite{AlonPPRS05, HatamiHPTZ22} lower bound sign rank, but neither quantity is able to give lower bounds that are superlogarithmic in the matrix dimensions. Alon, Frankl, and R\"odl~\cite{AFR85} used a counting argument and tools from algebraic geometry to prove a nearly-maximal lower bound for random matrices; their argument was subsequently refined by~\cite{BenDavidES03, alon2014sign} to handle low VC classes. The first superlogarithmic lower bound on the sign rank of an explicit matrix was given by Forster~\cite{Forster02}, whose average margin technique spurred an extensive line of work understanding the sign rank of specific matrices and clarifying the relationship of sign rank to other Boolean complexity measures~\cite{forsteretal, forster2002smallest, linial2007complexity, sherstovsymm, RS10, BT16, ChattopadhyayM17, HatamiQ17, BCHTV17, BunMT21, SherstovW19, ChattopadhyayM22}. Finally,~\cite{HatamiHM23, BlondalHHLT26, FrickHV26, BlondalHHLT26b} used topological arguments to study the unbounded error communication complexity of the Gap Hamming Distance problem, and extended these to more general topological lower bound frameworks. The threads from prior work that are most closely related to this paper are those 1) lifting sign degree lower bounds to sign rank via Forster's average margin condition and 2) passing through a problem in continuous space to permit the use of topological arguments. We describe these connections in more detail below, as well as technical connections to prior work in learning theory.

\paragraph{Lifting Sign Degree to Sign Rank.} Unbounded error communication and sign rank have the following direct analog in query (a.k.a. decision tree) complexity: 
The \emph{sign degree} (a.k.a. threshold degree) of $f$, denoted $\sdeg(f)$, is the least degree of a real polynomial $p$ such that $\sgn(p(x)) = f(x)$ for every $x \in \{-1, 1\}^n$. The analog of the question ``Is $\UPPcc$ closed under intersection?'' for UPP query complexity and sign degree is now well-established to be answered in the negative~\cite{Sherstov13, Sherstov13b, Sherstov21}. The most dramatic separation from this line of work~\cite{Sherstov21} gives an explicit family of Boolean halfspaces $f: \{-1, 1\}^n \to \{-1, 1\}$ such that $\sdeg(f) = O(1)$ but $\sdeg(f \land f) = \Omega(n)$.

The conceptual analogy between UPP communication and query complexity can be made technical as well.
There has been remarkable success in proving query-to-communication lifting theorems that upgrade a lower bound for a problem in the (weaker) query model to a lower bound for a related problem in the (stronger) communication model. In the context of the unbounded error models, a dream lifting theorem would show that there is a constant-sized two-party gadget $g$ such that $U(f \circ g^n) = \Omega(\sdeg(f))$ for every Boolean function $f$. Or equivalently, that the sign rank of (the communication matrix associated to) $f \circ g^n$ is exponential in the sign degree $\sdeg(f)$.

While it is still unknown whether such a generic lifting theorem holds for UPP complexity (in particular, one would have already implied that $\UPPcc$ is not closed under intersection) important partial progress has been made in this direction. Much technical work on sign degree relies on its dual characterization: a function $f$ has sign degree greater than $d$ if and only if there is a distribution $\mu$ under which $f$ is orthogonal to all polynomials of degree $d$. Building on the pattern matrix method~\cite{Sherstov11}, Sherstov~\cite{sherstovsymm} and Razborov and Sherstov~\cite{RS10} showed that a sign degree lower bound indeed lifts to a sign rank lower bound if this dual distribution $\mu$ can be made \emph{smooth} in the sense of placing noticeable probability mass on most inputs. This smoothness is critical to enable transferring a spectral norm bound to a sign rank bound via Forster's method. A number of subsequent works proved new sign rank lower bounds by constructing these smooth dual distributions~\cite{BT16, BCHTV17, BunT21, SherstovW19, BunMT21}. In particular, the previous work of~\cite{BunMT21}, showing that sign rank can increase quasipolynomially under intersection, did so by exhibiting a smooth dual distribution for the intersection of two majorities.

Our technical approach can be thought of as carrying out an analogous program, though for our continuous problem on products of unit spheres rather than directly on the Boolean hypercube. Recall that our argument ultimately relies on the inapproximability of the function $Q_2(s_1, s_2) = \sgn s_1 \land \sgn s_2$ by low degree polynomials. Over the domain $([-1, -2^{-Cd}] \cup [2^{-Cd}, 1])^2$, this function has sign degree at least $d$. As in the discrete case, however, high sign degree alone seems insufficient to yield an average margin bound. We need to improve it (via \Cref{lem:poly-lb-informal} /~\Cref{lem:poly-lower-bound}) to show that low-degree polynomials that nearly sign-represent $Q_2$ must be uniformly bounded. While this stronger condition is stated purely from the ``primal'' point of view, it actually implies the existence of a smooth dual distribution (and in our view, is morally equivalent to one). In fact, the work of~\cite{sherstovsymm, RS10} in the discrete setting derived their smooth dual distributions as a consequence of similar primal statements (see also~\cite[Section 10.5]{bun-tha:approximate-degree} for further discussion of the relationship). Finally, our spherical averaging argument can be viewed as a continuous analog of ``lifting'' from smooth sign degree to sign rank: it shows that regardless of the choice of the arrangement $f, g$, their symmetrization $H$ behaves as if it were a low degree polynomial, while carrying the same margin information.

\paragraph{Topological Sign Rank Lower Bounds.} In the $(n, k)$-Gap Hamming Distance problem, Alice and Bob hold vectors $a, b \in \{-1, 1\}^n$ and their goal is to determine whether $d_{\text{Hamming}}(a, b) \le k$ or $d_{\text{Hamming}}(a, b) \ge n-k$ under the promise that one of these conditions holds. Hatami, Hosseini, and Meng~\cite{HatamiHM23} proved that the sign rank of this problem's communication matrix is $\Omega\left(\frac{k}{\log(n/k)}\right)$ through a short and elegant topological argument. Like us, they shifted their attention to a continuous analog of the problem where Alice and Bob instead hold inputs on the unit sphere $\mathbb{S}^{n-1}$. The Borsuk--Ulam Theorem from topology immediately rules out the existence of continuous linear arrangements $f, g : \mathbb{S}^{n-1} \to \mathbb{S}^{r-1}$ for any dimension $r < n$. An approximation argument extends this to hold for arbitrary low-dimensional arrangements, while a randomized reduction transfers the lower bound to the original discrete problem.

Subsequent work~\cite{FrickHV26} generalized this topological argument to handle input shapes beyond the unit sphere, improving the Gap Hamming sign rank lower bound to $\Omega(k)$. However, their ``$\mathbb{Z}_2$ -index'' framework was very recently shown to be unable to prove sign rank lower bounds above $\log^2 N$~\cite{FrickHMNV26} for $N \times N$ matrices, which correspond to doubly logarithmic UPP communication lower bounds. Meanwhile, studying Forster's average margin condition in continuous space permits us to prove quantitatively much larger lower bounds for our problem. Complementary work of Blondal et al.~\cite{BlondalHHLT26b} studies the relationship between the $\mathbb{Z}_2$ index and a stronger lower-bound parameter, the list replicability number.

\paragraph{Lower Bounds for Learning via Spherical Harmonics.} Recall that a key component of our technical approach is to use a rotational symmetrization argument to reduce the problem of bounding the average margin of $\CHS_d \land \CHS_d$ to a question of approximability by low-degree polynomials. While this spherical symmetrization appears to be new in the sign rank and communication complexity literature, similar approaches have been used before in related settings in learning theory.

Daniely, Linial, and Shalev-Shwartz~\cite{DanielyLS14} proved lower bounds for agnostically learning large-margin halfspaces via generalized linear methods. Their learning models constrain how a learning algorithm represents a hypothesis the same way as in dimension complexity / sign rank, i.e., as a halfspace over feature maps. However, they impose the additional requirement that the algorithm selects this hypothesis by minimizing a convex surrogate loss, and thus their lower bound techniques do not have direct consequences for dimension complexity.
Nevertheless, much of their approach foreshadowed our analytic argument. Their argument begins with a lower bound against learning one-dimensional threshold functions by minimizing a convex loss over low-degree polynomials. They then extend this lower bound to higher-dimensional unit spheres by exploiting a similar rotational averaging argument: A learner for the high-dimensional learning problem can be symmetrized to depend only on an inner product, and this symmetrization admits a low-degree polynomial approximation.

Subsequent work of Daniely~\cite{Daniely17} showed that if a function $f(x, y)$ on two spherical variables can be written as $f(x, y) = g(\langle x, y\rangle)$ where $g$ cannot be approximated by low-degree polynomials, then $f$ cannot be approximated by small depth-two neural networks with exponentially bounded weights. He proved this result using related machinery from spherical harmonics, with the major difference being that his goal was to prove lower bounds against $L_2$ approximation rather than sign representation.

\subsection{Open Questions} \label{sec:open}

\begin{itemize}
    \item Can one exhibit a function $F$ exhibiting a nearly maximal separation $U(F) = O(\log n)$ while $U(F \land F) = \Theta(n)$? \Cref{thm:HS-tight} shows that our lower bound analysis is optimal, so our construction does not achieve this. 
    \item Is in the case that $\sup_{M} \sr(\HS_{M, d} \land \HS_{M, d}) < \infty$ for every fixed dimension $d$? A negative answer would resolve the open question of~\cite{HatamiHPTZ22} on whether there exists a function $\zeta$ such that $U(F \land F) \le \zeta(U(F))$ for every function $F$. Note that such a proof would have to go beyond average margin lower bounds, as the average margin of an intersection of two matrices is at most exponential in their sign rank~\cite[Corollary B.1]{HatamiHPTZ22}.
    \item In this work, we study the relationship between $U(F)$ and $U(G(F, \dots, F))$ where $G$ is the ``combining'' function $\AND_2$ (or, more generally, $\AND_k$ for constant $k$). How much larger can $U(G(F, \dots, F))$ be than $U(F)$ in terms of $G$?
    \item Do our techniques point to a more general lifting theorem from sign degree to sign rank that can complement the existing line of work based on the pattern matrix method? In particular, are there sign degree dual witnesses whose smoothness does not suffice to apply existing pattern-matrix results, but could be lifted using our techniques?
\end{itemize}

\paragraph{AI Acknowledgment.} This paper arose out of our prompting OpenAI's GPT-6 Astra model for ideas on how to improve the separation of~\cite{BunMT21} beyond the quadratic gap they achieved. To our (in retrospect, na\"ive) astonishment, it discovered what appeared to be complete proofs that (slightly different parameterizations) of the functions we consider in this paper yield the separation $U(f) = O(\log n)$ vs. $U(f \land f) = \Omega(n^{1/4})$. Subsequent interactions with GPT-6 Astra helped us understand the argument, simplify it, and improve it quantitatively. The text of this manuscript was written by the authors, with additional AI assistance used to locate references and provide feedback on our exposition. The use of AI tools does not affect our expected responsibilities as authors. We thank OpenAI for access to these tools through their ChatGPT for Academic Researchers program.

\paragraph{Paper Roadmap.} Section~\ref{sec:spherical-avg} gives the details of our symmetrization argument mapping a linear arrangement $f, g$ in continuous space to a function $H$ of inner products. Its main purpose is to establish ``property (c)'' from our technical overview, that this function $H$ has bounded derivatives and is approximated well by its Taylor polynomials. Section~\ref{sec:quadrant-bound} establishes \Cref{lem:poly-lb-informal} via \Cref{lem:poly-lower-bound}, controlling the coefficients of any real polynomial that nearly sign-represents the quadrant indicator function. In Section~\ref{sec:continuous-bound}, we complete the proof of our average margin upper bound for the continuous problem $\CHS_{d} \land \CHS_{d}$. Section~\ref{sec:rounding} describes a generic argument for transferring our continuous margin upper bound to margin upper bounds in the discrete setting, while Section~\ref{sec:main} instantiates these tools to prove our main results for bounded integer halfspaces and halfspaces defined by random points on the sphere. Section~\ref{sec:tightness} demonstrates the tightness of our margin and sign rank bounds. Finally, in Section~\ref{sec:applications}, we give the remaining details needed to establish our applications.

While our technical overview focused on studying the intersection of two disjoint copies of a problem, we choose to state many of our technical results in the more general setting of $k$-fold intersections. This choice, in our opinion, improves the legibility of our statements in these sections (where it might otherwise be easy to confuse the two parties Alice and Bob with the two copies of the problem they are trying to solve) and hints at how future work might generalize these techniques. Nevertheless, we warn the reader that despite larger values of $k$ yielding harder problems, some of our statements become (inconsequentially) \emph{weaker} in their ability to yield lower bounds with increasing $k$, as an artifact of how the proofs are organized. It will be of essentially no loss for the reader to mentally fix $k = 2$ as they read the rest of this paper.

\section{Preliminaries} \label{s:prelims}

We denote by $\N := \{1,2,\ldots\}$ the set of natural numbers, by $\Z$ the set of integers, by $\R$ the set of real numbers, and by $\mathbb{C}$ the set of complex numbers.
We write $X^{m \times n}$ for the set of $m \times n$ matrices with entries in $X$.
For finite sets $\mathcal{A}, \mathcal{B}$, we write $X^{\mathcal{A} \times \mathcal{B}}$ for the set of matrices with entries in $X$ whose rows and columns are indexed by elements of $\mathcal{A}$ and $\mathcal{B}$ respectively.
Throughout this paper, we identify $-1$ and 1 with the Boolean values ``true'' and ``false'' respectively.
We use $\land$ to denote Boolean AND, that is, $x \land y$ evaluates to $-1$ if $x = y = -1$ and to 1 otherwise.
For an event $P$, we write $\mathbbm{1}[P]$ for the indicator of $P$. We use $\mathbf{1}_n$ to be $n$-dimensional vector where each coordinate is 1; we suppress $n$ when it is clear from context.

Given a point $x \in \R^d$ define the \emph{$\ell_1$-norm} of $x$ as $\|x\|_1 := \sum_{i = 1}^d |x_i|$ and the \emph{$\ell_2$-norm} of $x$ as $\|x\|_2 := (\sum_{i = 1}^d x_i^2)^{1/2}$.
For $d \ge 2$, let $\mathbb{S}^{d-1} \subseteq \R^d$ denote the $(d-1)$-dimensional unit sphere in $d$ ambient dimensions. 
We use ``$\log$'' and ``$\ln$'' to denote the base-2 and natural logarithms respectively.
We use the notation $\exp(x)$ to denote $e^x$ where $e$ is Euler's number.
The \emph{sign} of a real number $x$, denoted $\sgn(x)$, is $-1$ if $x < 0$, $+1$ if $x > 0$, and 0 if $x = 0$.

A \emph{$k$-multi-index} is a finite ordered list of non-negative integers $\alpha = (\alpha_1, \ldots, \alpha_k)$.
In most of our applications the number of indices $k$ will be clear from context, so we will often suppress it.
We define $|\alpha| := \sum_{i=1}^k |\alpha_i|$ and $\alpha! := \prod_{i=1}^k \alpha_i!$.
Given a $k$-variable input $x = (x_1, \ldots, x_k)$ we use $x^\alpha$ to denote the monomial $\prod_{i = 1}^k x_i^{\alpha_i}$.
Thus an arbitrary $k$-variate polynomial $p$ of degree at most $D$ with real coefficients can be written as $p(x) = \sum_{|\alpha| \leq D} \lambda_\alpha x^\alpha$ for some $\lambda_\alpha \in \R$, where the sum runs over all $k$-multi-indices $\alpha$ with $|\alpha| \leq D$.
All polynomials in this paper have real coefficients.

Let $f, g : X \to \R$ be arbitrary functions.
We say $f$ \emph{sign-represents} $g$ if for every $x \in X$ we have $f(x) \cdot g(x) > 0$.
Similarly, given matrices $M_1, M_2 \in \R^{\mathcal{A} \times \mathcal{B}}$, we say $M_1$ sign-represents $M_2$ if $M_1(a,b) \cdot M_2(a,b) > 0$ for every $(a, b) \in \mathcal{A} \times \mathcal{B}$.
The \emph{sign degree} (a.k.a.\ \emph{threshold degree}) of a Boolean function $f : \pmone^n \to \pmone$, denoted $\deg_\pm(f)$, is the minimum degree of a polynomial $p : \pmone^n \to \R$ that sign-represents $f$.
The \emph{sign rank} of a matrix $M \in \pmone^{\mathcal{A} \times \mathcal{B}}$, denoted $\sr(M)$, is the minimum rank of a matrix $\widetilde{M} \in \R^{\mathcal{A} \times \mathcal{B}}$ that sign-represents $M$.

The communication complexity of a two-party function $F : \pmone^n \times \pmone^n \to \pmone$ is closely connected to properties of its \emph{communication matrix} $M_F \in \pmone^{2^n \times 2^n}$ defined by $M_F(x,y) = F(x,y)$.
Recall from the introduction that the \emph{unbounded-error communication complexity} of $F$, denoted $U(F)$, is the minimum number of bits communicated in a two-party communication protocol that computes $F$ on each input with probability strictly greater than $1/2$.
Paturi and Simon~\cite{PS86} showed the following equivalence between $U(F)$ and the sign rank of $M_F$.

\begin{theorem}\label{thm: ps}
Let $F : \pmone^{n} \times \pmone^n \to \pmone$ be a function and let $M_F$ be its communication matrix.
Then $U(F) = \log (\sr(M_F)) \pm O(1)$.
\end{theorem}

We will often refer to the function $F$ and its communication matrix $M_F$ interchangeably, e.g., we will write $\sr(F)$ to mean $\sr(M_F)$.

For $\eta \in (0,1)$, we define $I_\eta := [-1, -\eta] \cup [\eta, 1]$.
Define $Q_k : \R^k \to \R$ as
\[
    Q_k(s_1, s_2, \ldots, s_k) := \bigwedge_{i = 1}^k \sgn s_i
\]
if $s_i \neq 0$ for each $i$, or 0 otherwise.
That is, $Q_k$ is the Boolean indicator of the ``lower left'' orthant in $\R^k$.

\subsection{Random Points on the Sphere}

We will need the following concentration inequalities about the inner product of two random points on the sphere. We include a proof in Appendix~\ref{app:tail-bounds}.
\begin{proposition}[store=tail-bounds] \label{lem:score-conc-bounds}
    Let $X$ and $Y$ be distributed independently and uniformly over $\mathbb{S}^{d-1}$. Let $S = \langle X, Y \rangle$.
    Then for every $\delta \in (0, 1)$,
    \begin{gather}
        \frac{1}{4} (1-\delta^2)^{d-1} \le \Pr[|S| \geq \delta] \leq \frac{1}{\delta} (1 + \delta^2)^{-(d-1)/2}, \text{ and} \label{eq:score-upper-tail} \\
        \Pr[|S| \leq \delta] \leq 2 \delta \sqrt{d}. \label{eq:score-lower-tail}
    \end{gather}
\end{proposition}

We will also use the fact that small neighborhoods of random points on a sphere cover the sphere with high probability.
The proof of this fact follows from standard probability estimates (see, e.g., the proof of \cite[Lemma~3.2]{Kobos23}).

\begin{lemma} \label{lem:delta-net}
    For every constant $a > 0$, there exists a constant $b > 0$ for which the following holds. For all sufficiently large $d$ and $N \geq e^{bd^2}$, the probability over independent uniformly random $X_1, \ldots, X_N \in \mathbb{S}^{d-1}$ that there exists $y \in \mathbb{S}^{d-1}$ such that for all $i \in [N]$, $\|y - X_i\|_2 > e^{-ad}$, is at most $1/N$.
\end{lemma}

\section{Derivatives of Spherical Averages}
\label{sec:spherical-avg}

Let $f, g : (\mathbb{S}^{d-1})^k \to \R^r$ be measurable functions such that $\|f(x)\|_2, \|g(x)\|_2 \le 1$ for every $x \in (\mathbb{S}^{d-1})^k$. We define a ``symmetrized'' function $H_{f, g}: [-1, 1]^k \to [-1, 1]$ such that $H_{f, g}(s_1, \dots, s_k)$ captures the average value of $\langle f(x_1, \dots, x_k), g(y_1, \dots, y_k)\rangle$ over all points for which $s_1 = \langle x_1, y_1\rangle, \dots, s_k = \langle x_k, y_k\rangle$. Formally, we define $H_{f, g}(s_1, \dots, s_k)$ as the expectation $\mathbb{E}_{X, Y}[\langle f(X), g(Y)\rangle]$ for random variables $X = (X_1, \dots, X_k), Y = (Y_1, \dots, Y_k)$ sampled according to the following experiment that we denote by $\mathcal{D}_s$:
\begin{enumerate}
\item Sample $X_1, \dots, X_k$ independently and uniformly from $\mathbb{S}^{d-1}$.
\item For each $i = 1, \dots, k$, sample $V_i$ independently and uniformly from the orthogonal complement $\{v \in \mathbb{S}^{d-1} \mid \langle v, X_i \rangle = 0\}$ of $X_i$ in $\mathbb{S}^{d-1}$.
\item For each $i = 1, \dots, k$, set $Y_i = s_i X_i + \sqrt{1 - s_i^2} \ V_i$.
\end{enumerate}
Each of $X$ and $Y$ are marginally distributed as $k$ independent points from the unit sphere, while satisfying the inner product conditions $s_i = \langle X_i, Y_i\rangle$ for every $i$.

When it is clear from context, we suppress the subscripts $f, g$ and write $H = H_{f, g}$. The goal of this section is to show that when $f, g$ are arbitrary bounded, measurable functions, their symmetrization $H$ has small derivatives near the origin, and is hence well-approximated by its Taylor polynomial.

\begin{theorem} \label{thm:sym-derivatives}
    Let $f, g : (\mathbb{S}^{d-1})^k \to \R^r$ be measurable with $\|f(x)\|_2, \|g(x)\|_2 \le 1$ for every $x$. The function $H = H_{f, g}$ is continuous on $[-1, 1]^k$ and continuously differentiable on $(-1,1)^k$ through order $\lfloor (d-2) / 2\rfloor$. Furthermore, for every multi-index $\alpha = (\alpha_1, \dots, \alpha_k)$ with $|\alpha| \le \lfloor (d-2) / 2 \rfloor$, the partial derivative $\partial^\alpha H = \partial^{\alpha_1}_1\partial^{\alpha_2}_2\dots\partial^{\alpha_k}_kH$ satisfies
    \[
        \frac{|\partial^\alpha H(s)|}{\alpha!} \le 4^{|\alpha|}
    \]
    for every $s \in [-1/4, 1/4]^k$.
\end{theorem}

The bulk of this section is devoted to the proof of~\Cref{thm:sym-derivatives}, and it may be skipped on first reading without loss of continuity. It uses standard ideas from spherical harmonics and builds on an approach to proving smoothness for ``isotropic positive definite'' functions from that literature~\cite{Ziegel14, TrubnerZiegel17}. We require a more quantitative treatment for our problem and have tried to make our presentation as self-contained as possible. The main idea is to expand the function $H$ as an infinite linear combination of polynomials. It then suffices to show that 1)  each individual polynomial in this expansion has uniformly bounded derivatives (\Cref{thm:gegenbound}) and 2) the total magnitude of the coefficients in this expansion is bounded (\Cref{thm:sym-series}). The polynomials suited for this task are products of Gegenbauer polynomials, which may be thought of as spherical averages of the Fourier frequencies on the unit sphere. We develop all of the machinery we need for these in Section~\ref{sec:gegenbauer}.

The following corollary estimating the Taylor remainder for $H$ is the only result from this section that is needed in the rest of this paper.

\begin{corollary} \label{cor:sym-taylor}
    Let $f, g$, $H = H_{f, g}$ be as in Theorem~\ref{thm:sym-derivatives}. For $0 < D \leq \lfloor (d-2) / 2\rfloor - 1$, let $T_D$ denote the Taylor polynomial of $H$ centered at the origin consisting of terms of total degree at most $D$. Then for every $s \in [-1/4, 1/4]^k$, we have
    \[|H(s) - T_D(s)| \le (4\|s\|_1)^{D+1}.\]
\end{corollary}

\begin{proof}[Proof of \Cref{cor:sym-taylor} assuming \Cref{thm:sym-derivatives}]

Fix $D \leq \lfloor (d-2) / 2\rfloor - 1$. For each $s \in [-1/4, 1/4]^k$, define the univariate function $h(\lambda) = H(\lambda s)$ for $\lambda \in [0, 1]$. Then $h(1) = H(s)$ and repeated application of the chain rule shows
\[h^{(j)}(0) = \sum_{|\alpha| = j} \frac{j!}{\alpha!} s_1^{\alpha_1} \dots s_k^{\alpha_k} \partial^\alpha H(0).\]
Hence
\[\sum_{j=0}^D \frac{h^{(j)}(0)}{j!} = \sum_{|\alpha| \le D} \frac{\partial^\alpha H(0)}{\alpha!}s_1^{\alpha_1} \dots s_k^{\alpha_k} = T_D(s).\]
The left-hand-side is simply the degree-$D$ Taylor polynomial of $h$ evaluated at $1$. So we can use the single-variable version of Taylor's theorem on the function $h$ to obtain
\[|H(s) - T_D(s)| = \left|h(1) - \sum_{j=0}^D \frac{h^{(j)}(0)}{j!} \right| \le \sup_{0\le \lambda \le 1} \frac{|h^{(D+1)}(\lambda)|}{(D+1)!}.\]
Now by the chain rule and~\Cref{thm:sym-derivatives},
\[\frac{|h^{(D+1)}(\lambda)|}{(D+1)!} \le \sum_{|\alpha| = D+1} |s_1|^{\alpha_1} \dots |s_k|^{\alpha_k} \cdot \frac{|\partial^\alpha H(\lambda s)|}{\alpha!} \le (4\|s\|_1)^{D+1}\]
for every $\lambda \in [0, 1]$, completing the proof.
    
\end{proof}

\subsection{Gegenbauer Polynomials} \label{sec:gegenbauer}

For a ``dimension'' parameter $d \ge 2$, define the weight function $v_d(t) = (1-t^2)^{(d-3)/2}$ on the interval $[-1, 1]$. Up to a constant factor, this function arises as the probability density of the inner product between two random vectors on the unit sphere $\mathbb{S}^{d-1}$. The \emph{Gegenbauer (a.k.a.\ ultraspherical) polynomials} $G_0^{d}, G_1^{d}, G_2^{d}, \dots$ form a basis for the vector space of real polynomials over $[-1, 1]$ that is orthogonal with respect to the inner product $\langle p, q \rangle_{v_d} := \int_{-1}^1 p(t) q(t) v_d(t) \, \dd t$. They may be constructed by applying the Gram--Schmidt orthogonalization procedure to the sequence of monomials $1, t, t^2, \dots$. That is, one obtains the degree-$i$ Gegenbauer polynomial $G_i^{d}$ as the (normalized) component of the monomial $t^i$ which is orthogonal to all of $G_0^{d}, G_1^{d}, \dots, G_{i-1}^{d}$ under the inner product $\langle \cdot, \cdot \rangle_{v_d}$. It will be convenient for us to normalize these polynomials to take the value $1$ at $t = 1$, which we capture in the following definition.

\begin{definition}
    For $d \ge 2$, the dimension-$d$ Gegenbauer polynomials $G_0^{d}, G_1^{d}, G_2^{d}, \dots$ comprise the unique sequence of polynomials on $[-1, 1]$ such that
    \begin{enumerate}
        \item Each $G_i^{d}$ has degree $i$ with a nonzero coefficient on $t^i$,
        \item For every $i \ne j$, we have 
                \[\int_{-1}^1 G_i^{d}(t) G_j^{d}(t) v_d(t) \, \dd t = 0\]
            where $v_d(t) = (1-t^2)^{(d-3)/2}$,
        \item For every $i$, we have $G_i^{d}(1) = 1$.
    \end{enumerate}
\end{definition}

We begin by recording some useful facts about these polynomials. Their proofs appear in Appendix~\ref{app:gegenbauer}.

\begin{proposition}[store=geg-facts]
    \label{prop:gegenbauer-facts}
    For every $d \ge 2$, the normalized Gegenbauer polynomials $G_i^{d}$ have the following properties.
    \begin{enumerate}
        \item \textbf{(Anti-)Symmetry.} For every $i$ and $t \in [-1, 1]$,
        \[G_i^{d}(-t) = (-1)^i G_i^{d}(t).\]
        \item \textbf{Spherical averaging identity.}
        Let $q : \mathbb{S}^{d-1} \to \mathbb{R}$ be a degree-$i$ polynomial such that $\mathbb{E}_{Y \in \mathbb{S}^{d-1}} [q(Y) r(Y)] = 0$ for every degree-$(i-1)$ polynomial $r$. Then for every $x \in \mathbb{S}^{d-1}$,
        \[P_x(t) := \mathop{\mathbb{E}}_{U \in \mathbb{S}^{d-1} : \langle U, x \rangle = 0}\left[q(tx + \sqrt{1-t^2}U)\right] = q(x) \cdot G_{i}^{d}(t).\]
        (This identity is essentially equivalent to the Funk-Hecke formula in spherical harmonics, e.g., ~\cite[Theorem 1.2.9]{DaiXu13}.)
        \item \textbf{Boundedness.} For every $i \ge 0$ and every $t \in [-1, 1]$, we have $|G_i^{d}(t)| \le 1$.
        \item \textbf{Dimension reduction~\cite[Equation (3.42)]{AskeyF69}.} For every $d \ge 4$ and every $i$, the polynomial $G_i^{d}$ is a convex combination of the polynomials $G_j^{d-2}$ where $0 \le j \le i$ and $j \equiv i \pmod 2$.
        \item \textbf{Derivative recurrence~\cite[special case of Proposition 4.4]{Ziegel14}.} For $d \ge 4$ and $i \ge 0$ we have
        \[(1-t^2) \frac{\dd}{\dd t} G_i^{d}(t) = (d-3)\left( tG_{i}^{d}(t) - G_{i+1}^{(d-2)}(t) \right).\]
    \end{enumerate}
\end{proposition}

The key estimate we need on the Gegenbauer polynomials is that they have uniformly bounded derivatives on subintervals of $[-1, 1]$. Our proof follows the dimension-changing approach of~\cite{Ziegel14, TrubnerZiegel17}.

\begin{theorem}\label{thm:gegenbound}
    Let $\rho \in [0, 1)$. For every order of differentiation $D \ge 0$, every dimension $d \ge 2D + 2$, and every $i \ge 0$, we have
    \[\max_{t \in [-\rho, \rho]} \frac{|(G_i^{d})^{(D)}(t)|}{D!} \le \left(\frac{3}{1-\rho}\right)^D.\]
\end{theorem}

\begin{proof}
    We proceed by induction on $D$. The base case of $D = 0$ is the boundedness condition Item 3 of Proposition~\ref{prop:gegenbauer-facts}. Let $D \ge 1$ and suppose the claim is true for all smaller derivative orders. By repeatedly applying Item~4 of Proposition~\ref{prop:gegenbauer-facts}, it suffices to prove the claim when $d \in \{2D+2, 2D+3\}$, depending on whether the original $d$ was even or odd.

    To simplify notation, let $P = G_{i}^{d}$ and let $R = G_{i+1}^{d-2}$. Define 
    \[a(t) = \frac{t}{1-t^2} \, \text{ and } \, b(t) = \frac{1}{1-t^2}.\]
    Then Item~5 of Proposition~\ref{prop:gegenbauer-facts} gives us the recurrence $P' = (d-3)(aP - bR)$. Differentiating a further $D-1$ times and applying the product rule gives
    \begin{equation} \label{eq:product-rule}
        P^{(D)} = (d-3) \sum_{j = 0}^{D-1} \binom{D-1}{j} \left( a^{(D-j-1)}P^{(j)} - b^{(D-j-1)} R^{(j)}\right).
    \end{equation}
    Observe that both $P$ and $R$ meet the conditions of the inductive hypothesis for all derivative orders $j \in \{0, 1, \dots, D-1\}$, as their dimensions are at least $d-2 \ge 2D \ge 2j+2$. Hence, $|P^{(j)}|, |R^{(j)}| \le j! (3/(1-\rho))^j$ on the interval $[-\rho, \rho]$. Meanwhile, the derivatives of $a$ and $b$ satisfy
    \begin{align*}
        \frac{\dd^{j}}{\dd t^j} (b(t) + a(t)) &= \frac{\dd^{j}}{\dd t^j} \frac{1}{1-t} = \frac{j!}{(1-t)^{j+1}}, \\
        \frac{\dd^{j}}{\dd t^j} (b(t) - a(t)) &= \frac{\dd^{j}}{\dd t^j} \frac{1}{1+t} = \frac{(-1)^j j!}{(1+t)^{j+1}},
    \end{align*}
    and hence 
    \[|a^{(j)}(t)| + |b^{(j)}(t)| = \max\{ |b^{(j)}(t) + a^{(j)}(t)|, |b^{(j)}(t) - a^{(j)}(t)|\} = \frac{j!}{(1-|t|)^{j+1}}.\]
    Applying these bounds to \eqref{eq:product-rule} and dividing by $D!$ gives, for every $t \in [-\rho, \rho]$,
    \begin{align*}
        \frac{|P^{(D)}(t)|}{D!} &\le \frac{d-3}{D!} \sum_{j=0}^{D-1} \binom{D-1}{j} \frac{(D-j-1)!}{(1-\rho)^{D-j}} \cdot j! \left(\frac{3}{1-\rho}\right)^j \\
        &= \frac{d-3}{D (1-\rho)^D} \cdot \sum_{j=0}^{D-1} 3^j \\
        &\le \frac{2}{(1-\rho)^D} \cdot \frac{3^D - 1}{2} \le \left(\frac{3}{1-\rho}\right)^D.
    \end{align*}
\end{proof}

\subsection{Series Expansions of Symmetrized Functions} \label{sec:series-sym}

In this section, we describe how to express the spherical symmetrization $H$ of any two bounded, measurable functions as an infinite series of products of Gegenbauer polynomials.     Let $p_0, p_1, p_2, \dots$ be the orthonormal basis for the space of real polynomials on $\mathbb{S}^{d-1}$ under the inner product $\langle p, q \rangle = \mathbb{E}_{x \sim \mathbb{S}^{d-1}}[p(x) q(x)]$ obtained by applying Gram--Schmidt to monomials in nondecreasing order of total degree. This ensures that every polynomial in this basis is orthogonal to all other polynomials of lower degree. Then $k$-fold products of these polynomials comprise an orthonormal basis for the space of polynomials on $(\mathbb{S}^{d-1})^k$. By the Stone--Weierstrass theorem, these polynomials are dense in the space of continuous functions on $(\mathbb{S}^{d-1})^k$, and hence also in the space of square-integrable functions, which we denote by $L_2 = L_2((\mathbb{S}^{d-1})^k)$. Write these basis elements as
\[\Psi_J(x_1, \dots, x_k) = \prod_{i=1}^k p_{j_i}(x_i),\]
where $J = (j_1, \dots, j_k) \in \{0, 1, 2, \dots\}^k$.

\begin{lemma}\label{lem:basis-correlation}
    Let $J = (j_1, \dots, j_k)$, $J' = (j_1', \dots, j_k')$, and $s \in [-1, 1]^k$. For each index $i \in [k]$, let $\ell_i = \deg(p_{j_i})$. Then
    \[
\mathbb E_{(X, Y) \sim \mathcal{D}_s}[\Psi_J(X)\Psi_{J'}(Y)]
=
\begin{cases}
\displaystyle\prod_{i=1}^kG_{\ell_i}^{d}(s_i) & \text{ if } J=J', \\
0 & \text{ if } J\ne J'.
\end{cases}
\]
\end{lemma}

\begin{proof}
First consider the case of a single sphere, i.e., $k=1$. The goal is to show that
\[\mathbb{E}[p_j(X) p_{j'}(Y)] = \begin{cases}
G_{\ell}^{d}(t) & \text{ if } j=j', \\
0 & \text{ if } j \ne j',
\end{cases}\]
where the expectation is taken over single spherical variables $X,Y$ with $\langle X, Y\rangle = t$ and $\ell = \deg(p_j) = \deg(p_{j'})$ in the case $j = j'$. The general claim will follow from the fact that the pairs $(X_i, Y_i)$ are independent across $i$ in the general $k$-sphere averaging experiment.

Item~2 of Proposition~\ref{prop:gegenbauer-facts} and the fact that $p_{j'}$ is orthogonal to all preceding polynomials allows us to write
\begin{align*}
    \mathbb{E}[p_j(X) p_{j'}(Y)] &= \mathbb{E}_{X \sim \mathbb{S}^{d-1}} \big[ p_j(X) \cdot \mathbb{E}_{V \sim \mathbb{S}^{d-1}: \langle V, X\rangle = 0}[p_{j'}(tX + \sqrt{1-t^2}V)] \big] \\
    &= \mathbb{E}_{X \sim \mathbb{S}^{d-1}} \big[ p_j(X) \cdot p_{j'}(X) \cdot G_{\deg(p_{j'})}^{d}(t) \big].
\end{align*}
The statement now follows by orthonormality.
\end{proof}

\begin{theorem} \label{thm:sym-series}
    Let $f, g : (\mathbb{S}^{d-1})^k \to \mathbb{R}^r$ be measurable functions with $\|f(x)\|_2, \|g(x)\|_2 \le 1$ for every $x$. Let $H : [-1, 1]^k \to [-1, 1]$ be their spherical symmetrization. Then
    \[H(s_1, \dots, s_k) = \sum_{\ell} c_{\ell} \prod_{i= 1}^k G^{d}_{\ell_i}(s_i)\]
    where the indices $\ell = (\ell_1, \dots, \ell_k)$ range over $\{0, 1, 2, \dots \}^k$ and the real coefficients $c_{\ell}$ satisfy
    \[\sum_{\ell} |c_\ell| \le 1.\]
    Moreover, the series converges uniformly on $[-1, 1]^k$.
\end{theorem}

\begin{proof}
Let $\{\Psi_J\}_{J \in \{0,1,2,\ldots\}^k}$ denote the orthonormal basis for the space of real polynomials on $(\mathbb{S}^{d-1})^k$ as described above. Then $f$ and $g$ may be written as
    \[f(x) = \sum_{J} a_{J} \Psi_J(x) \, \text{ and } \, g(x) = \sum_{J} b_{J} \Psi_J(x)\]
    for coefficients $a_J, b_J \in \mathbb{R}^r$.
    For a tuple of indices $J = (j_1, \dots, j_k)$, let $D(J) = (\deg(p_{j_1}), \dots, \deg(p_{j_k}))$. We will show that it suffices to take $c_\ell = \sum_{J : D(J) = \ell} \langle a_J, b_J \rangle$ for every tuple  of degrees $\ell$.

    First observe that by Cauchy--Schwarz and then by Parseval's identity applied to the orthonormal $L_2$ expansions of $f$ and $g$,
    \[\sum_{\ell} |c_\ell| \le \left(\sum_{J}\|a_J\|^2_2\right)^{1/2} \left(\sum_{J}\|b_J\|^2_2\right)^{1/2} = \|f\|_{L_2} \|g\|_{L_2} \le 1.\]
    
    We now show that the $c_\ell$'s yield a correct, uniformly convergent series representation of $H$. For each $m \ge 0$, consider the finite partial sums
    \[f_m(x) = \sum_{J : \|D(J)\|_\infty \le m} a_{J} \Psi_J(x), \qquad g_m(x) = \sum_{J : \|D(J)\|_\infty \le m} b_{J} \Psi_J(x).\]
    Then for every $s \in [-1, 1]^k$, we have
    \begin{align*}
        H_m(s_1, \dots, s_k) &:= \mathbb{E}_{(X, Y) \sim \mathcal{D}_s}[\langle f_m(X), g_m(Y)\rangle] \\
        &= \sum_{J, J': \|D(J)\|_\infty, \|D(J')\|_\infty \le m} \langle a_J , b_{J'}\rangle \mathbb{E}_{(X, Y) \sim \mathcal{D}_s}[\Psi_J(X) \Psi_{J'}(Y)] \\
        &= \sum_{\ell : \|\ell\|_\infty \le m} \sum_{J: D(J) = \ell} \langle a_J , b_J\rangle \prod_{i=1}^k  G_{\ell_i}^{d}(s_i).
    \end{align*}
Here, the last equality holds by Lemma~\ref{lem:basis-correlation}.
Hence, it suffices to show that $H_m$ converges to $H$ uniformly. To see this, we calculate
\begin{align*}
    |H(s) - H_m(s)| &= \big| \mathbb{E}_{(X, Y) \sim \mathcal{D}_s}[\langle f(X), g(Y)\rangle] - \mathbb{E}_{(X, Y) \sim \mathcal{D}_s}[\langle f_m(X), g_m(Y)\rangle] \big| \\
     &= \big| \mathbb{E}_{(X, Y) \sim \mathcal{D}_s}[\langle f(X) - f_m(X), g(Y)\rangle] + \mathbb{E}_{(X, Y) \sim \mathcal{D}_s}[\langle f_m(X), g(Y) - g_m(Y)\rangle] \big| \\
    &\le \|f-f_m\|_{L^2} \|g\|_{L^2} + \|f_m\|_{L^2} \|g - g_m\|_{L^2}.
\end{align*}
The last inequality follows from Cauchy--Schwarz and holds for every value of $s$. Hence convergence of $f_m$ to $f$ and $g_m$ to $g$ in $L^2$ imply that this tends to $0$ uniformly in $s$.
\end{proof}

\subsection{Proof of Theorem~\ref{thm:sym-derivatives}}

Recall that our goal is to show that for all bounded, measurable functions $f, g$ their symmetrization
\[H(s_1, \dots, s_k) = \mathbb{E}_{(X, Y) \sim \mathcal{D}_s}[\langle f(X), g(Y)\rangle]\]
has bounded derivatives for every $s \in [-1/4, 1/4]^k$. Using Theorem~\ref{thm:sym-series}, write
\[H(s_1, \dots, s_k) = \sum_{\ell} c_\ell \prod_{i=1}^k G_{\ell_i}^{d}(s_i)\]
as a uniformly convergent series with $\sum_{\ell} |c_\ell| \le 1$. This immediately implies $H$ is continuous. For each $\ell$ and multi-index $\alpha$, the differentiated term
\[\partial^\alpha c_\ell \prod_{i=1}^k G_{\ell_i}^{d}(s_i) = c_\ell \prod_{i=1}^k (G_{\ell_i}^{d})^{(\alpha_i)}(s_i)\]
is, by Theorem~\ref{thm:gegenbound}, at most $|c_\ell| \cdot \alpha_1! \dots \alpha_k! \cdot \left(\frac{3}{1-\rho}\right)^{|\alpha|}$ in absolute value for every $s \in [-\rho, \rho]^k$. Then again since $\sum_{\ell} |c_\ell| \le 1$, each infinite series comprised of these terms converges uniformly on $[-\rho, \rho]^k$ for every $\rho \in (0, 1)$, so the expression
\[\partial^\alpha H(s) = \sum_{\ell} c_\ell \prod_{i=1}^k (G_{\ell_i}^{d})^{(\alpha_i)}(s_i)\]
is correct on $(-1, 1)^k$.
Applying the termwise upper bound of Theorem~\ref{thm:gegenbound} with $\rho = 1/4$ concludes the proof.
\qed

\section{Bounds for Approximations to the Orthant} \label{sec:quadrant-bound}

Recall the definition of $Q_k$ (for element-wise non-zero inputs).
\[
    Q_k(s_1, s_2, \ldots, s_k) = \bigwedge_{i = 1}^k \sgn s_i.
\]
In this section we show that if a low-degree real polynomial nearly sign-represents $Q_k$ except possibly on narrow bands around the coordinate hyperplanes, then it must have small coefficients.

\begin{lemma} \label{lem:poly-lower-bound}
    There exist universal constants $c, C > 0$ such that the following is true.
    Let $k \geq 2$ and $D \geq 1$ be integers, and set $\eta = e^{-D/C}$. Let $p(s_1, \ldots, s_k) = \sum_{|\alpha| \leq D} \lambda_{\alpha} s^\alpha$ be a $k$-variate real polynomial of degree at most $D$. If for all $s \in I_{\eta}^k$,
    \[Q_k(s) p(s) \geq -1,\]
    then for all $\alpha$, 
    \[|\lambda_{\alpha}| \leq e^{cD^2 + D \ln(40k)}.\]
\end{lemma}

In Section~\ref{sec:strict-lb}, we begin by proving the weaker statement that no polynomial of degree $\approx D$ strictly sign represents $Q_k$ on all of $I_\eta^k$. Section~\ref{sec:coefficient-lb} then shows how to upgrade this lower bound to show that even approximate sign representations must have small coefficients.

\subsection{Degree Lower Bounds for Strict Sign Representations} \label{sec:strict-lb}

The following lemma is implicit in Newman's proof that $|x|$ cannot be approximated pointwise by low-degree rational functions.
A complete proof appears in Appendix~\ref{app:newman}.

\begin{lemma}[store=newman,
note={Implicit in \cite{newman1964approximation}},
restate-keys={note={}}]
\label{lem:newman}
    Let $\eta \in (0,1)$ and $k \geq 2$ be an integer.
    If $p, q$ are real polynomials of degree at most $D$ such that for all $x \in I_\eta$, $|\frac{p(x)}{q(x)} - \sgn x| \leq 1/k$, then $D \geq 2C\ln(2k-1) \ln(1/\eta)$ where $C > 0$ is a universal constant.
\end{lemma}

We use a theorem due to Sherstov that shows that a sign representation for the conjunction of several functions can be used to obtain rational approximations for the individual functions whose total error across the functions is bounded.

\begin{theorem}[\!\cite[Theorem 3.17]{Sherstov13}] \label{thm:she-composition}
    Let $f_1, \ldots, f_k$ be Boolean functions on nonempty compact sets $\mathcal{X}_1, \ldots, \mathcal{X}_k \subseteq \R$ respectively.%
    \footnote{While Sherstov states his theorem for finite $\mathcal{X}_i$, his result extends to infinite compact sets. See the remark on page~15 of \cite{Sherstov13}.}
    Assume that $f_i \not\equiv 1$ for all $i \in [k]$, and let $D = \deg_\pm(\bigwedge_{i \in [k]} f_i)$.
    Then there exist polynomials $p_1, \ldots, p_k$ and $q_1, \ldots, q_k$ of degree at most $12D\ln(2k)$ such that $q_i(x) > 0$ for all $i$ and $x \in \mathcal{X}_i$, and
    \[
        \sum_{i=1}^k \left( \sup_{x \in \mathcal{X}_i} \left| f_i(x) - \frac{p_i(x)}{q_i(x)} \right| \right) < 1.
    \]
\end{theorem}

\begin{lemma} \label{lem:strict-sign-lb}
    There is a universal constant $C > 0$ such that the following holds.
    Let $\eta \in (0,1)$ and $k \geq 2$.
    If a $k$-variate real polynomial $q$ of degree $D$ sign-represents $Q_k$ on $I_\eta^k$, then $D > 2C\ln(1/\eta)$.
    Consequently, if $\eta = 2e^{-D/C}$ then no polynomial of degree $D$ sign-represents $Q_k$ on $I_{\eta}^k$.
\end{lemma}
\begin{proof}
    Let $C' > 0$ be the constant from \Cref{lem:newman}.
    For $i \in [k]$ let $\mathcal{X}_i = I_\eta$ and $f_i = \sgn|_{\mathcal{X}_i}$.
    Notice that $Q_k|_{I_\eta^k} = \bigwedge_{i = 1}^k f_i$, and so $q$ sign-represents $\bigwedge_{i = 1}^k f_i$ on its domain.
    As we will prove a lower bound on $D$, it is without loss of generality to assume that $D = \deg_\pm(\bigwedge_{i=1}^k f_i)$.
    By \Cref{thm:she-composition}, there exist polynomials $p_1, \ldots, p_k$ and $q_1, \ldots, q_k$ of degree at most $12D\ln(2k)$ such that
    \[
        \sum_{i=1}^k \left( \sup_{x \in I_\eta} \left| \sgn(x) - \frac{p_i(x)}{q_i(x)} \right| \right) < 1.
    \]
    Letting $\widetilde{p}, \widetilde{q}$ be the polynomials $p_i, q_i$ that minimize the summand, we get that $|\frac{\widetilde{p}(x)}{\widetilde{q}(x)} - \sgn x | < 1/k$ for all $x \in I_\eta$.
    \Cref{lem:newman} implies $12D\ln(2k) \geq C' \ln((2k-1)^2) \ln(1/\eta)$ and hence $D > \frac{C'}{12} \ln(1/\eta) = 2C \ln(1/\eta)$ where we set $C = C'/24$.
    
    If $\eta = 2e^{-D/C}$ then we get that every polynomial that sign-represents $Q_k$ on $I_{\eta}^k$ must have degree strictly greater than $2C (D/ C - \ln(2)) \geq 2C \cdot \frac{D}{2C} = D$.
\end{proof}

\subsection{Coefficient Upper Bounds for Approximate Sign Representations} \label{sec:coefficient-lb}

To convert the degree lower bound for exact sign representations to a coefficient upper bound for approximate sign representations, we show that the coefficients of a degree-$D$ polynomial in $k$ variables are bounded in magnitude by the integral of the magnitude of the polynomial over an appropriate interval.
We use facts about a certain sequence of polynomials which we first record.

Denote by $L_0, L_1, \ldots$ the polynomials
\[
    L_i(t) = \sum_{r=0}^i \binom{i}{r}^2 (t-1)^r (t-2)^{i-r}.
\]
For readers who are familiar with Legendre polynomials, $L_i(t)$ is exactly $P_i(2t-3)$ where $P_i$ is the standard degree-$i$ Legendre polynomial normalized with $P_i(1) = 1$.

\begin{proposition}[store=legendre-props] \label{prop:legendre} ~
    \begin{enumerate}
        \item For all $i, j$,
        \[
            \int_1^2 L_i(t) L_j(t) \, \dd t = \begin{cases}
                \frac{1}{2i+1} & \text{if } i = j \\
                0 & \text{otherwise.}
            \end{cases}
        \]
        \item For all $i$ and $t \in [1,2]$, $|L_i(t)| \leq 1$.
        \item The sum of the magnitudes of the coefficients of $L_i$ is at most $12^i$.
    \end{enumerate}
\end{proposition}

\Cref{prop:legendre} follows from several standard facts about Legendre polynomials (see, e.g.,~\cite[Sections 18.3, 18.5, 18.14]{DLMF}).
For completeness, we include a self-contained proof in Appendix~\ref{app:legendre}.

We can use \Cref{prop:legendre} to construct a basis for the set of $k$-variate polynomials of degree at most $D$.
We can then write an arbitrary $k$-variate polynomial $p$ with degree $D$ as a linear combination of these basis polynomials.
Applying the bounds on the $L_i$'s from the proposition, we get the following upper bound on the magnitudes of the coefficients of $p$ in terms of the integral of $|p|$ over an interval.

\begin{lemma} \label{lem:coeff-int-ub}
    For all $k, D \geq 1$, if $p(s_1, \ldots, s_k) = \sum_{|\alpha| \leq D} \lambda_\alpha s^\alpha$ is a $k$-variate polynomial of degree at most $D$, then for all $\alpha$,
    \[
        |\lambda_\alpha| \leq (40k)^D \int_{[1,2]^k} |p(s)| \,\dd s.
    \]
\end{lemma}
\begin{proof}
    For a multi-index $\xi = (\xi_1, \ldots, \xi_k)$, define
    \[
        L_\xi(s)=\prod_{i=1}^k L_{\xi_i}(s_i).
    \]
    Item~1 of \Cref{prop:legendre} implies that for all multi-indices $\xi, \theta$,
    \[
        \int_{[1,2]^k} L_{\xi}(s) L_{\theta}(s) \,\dd s = \begin{cases}
            \prod_{i = 1}^k \frac{1}{2\xi_i + 1} & \text{if } \xi = \theta \\
            0 & \text{otherwise.}
        \end{cases}
    \]
    So the $L_\xi$'s with $|\xi| \leq D$ form a basis for $k$-variate polynomials with degree at most $D$.

    Let $p(s_1, \ldots, s_k) = \sum_{|\alpha| \leq D} \lambda_\alpha s^\alpha$ be an arbitrary $k$-variate polynomial of degree at most $D$.
    Since the $L_\xi$'s form a basis, we can write
    \[
        p(s) = \sum_{|\xi|\leq D} \lambda'_{\xi} L_{\xi}(s).
    \]
    For any $\theta$ with $|\theta| \leq D$, multiplying both sides by $L_\theta(s)$ and integrating gives
    \begin{align*}
        \int_{[1,2]^k} p(s) L_\theta(s) \,\dd s &= \sum_{|\xi| \leq D} \lambda'_{\xi} \int_{[1,2]^k} L_{\xi}(s) L_\theta(s) \,\dd s \\
        &= \lambda'_\theta \prod_{i=1}^k \frac{1}{2\theta_i + 1}
    \end{align*}
    Rearranging yields
    \[\lambda'_\theta = \left( \prod_{i=1}^k (2\theta_i + 1) \right)
          \int_{[1,2]^k} p(s) L_\theta(s) \,\dd s.\]
    By item~2 of \Cref{prop:legendre}, $|L_\theta(s)|\leq 1$ on $[1,2]^k$.
    So applying the triangle inequality,
    \[
        |\lambda'_\theta|
        \leq \left( \prod_{i=1}^k (2\theta_i+1) \right) \int_{[1,2]^k} |p(s)|\,\dd s
        \leq 3^{|\theta|} \int_{[1,2]^k} |p(s)|\,\dd s,
    \]
    where the second inequality holds due to the fact that $2n + 1 \leq 3^n$ for every non-negative integer $n$.
    Let $m_\xi$ be the sum of the magnitudes of all coefficients of $L_\xi$. Item~3 of \Cref{prop:legendre} implies $m_\xi \leq \prod_{i=1}^k m_{\xi_i} \leq 12^{|\xi|}$.
    Combining these bounds and using the triangle inequality,
    we obtain
    \begin{align*}
        \max_{\alpha} |\lambda_\alpha|
        &\leq \sum_{|\xi| \leq D}
               |\lambda'_\xi| \cdot m_\xi \\
        &\leq \left( \sum_{|\xi|\leq D} 36^{|\xi|} \right) \int_{[1,2]^k} |p(x)|\,\dd x.
    \end{align*}

    For every $\ell \geq 0$, there are at most $k^\ell$ multi-indices $\xi$ with $|\xi| = \ell$.
    Therefore
    \[
        \sum_{|\xi|\leq D} 36^{|\xi|}
        \leq \sum_{\ell = 0}^D (36k)^\ell
        \leq \sum_{\ell = 0}^D \binom{D}{\ell} (36k)^\ell
        = (1+36k)^D
        \leq (40k)^D.
    \]
\end{proof}

We are ready to prove \Cref{lem:poly-lower-bound}.
The proof shows that if a low-degree approximate sign representation $p$ has a large coefficient, then it can be converted to a low-degree strict sign representation $q$, contradicting \Cref{lem:strict-sign-lb}.
Roughly speaking, $q$ is defined so that $Q_k(s)q(s)$ is bounded below by the integral of $|p|$ minus a constant.
\Cref{lem:coeff-int-ub} tells us that this quantity is bounded below by the largest coefficient of $p$ minus a constant, which is positive: $q$ sign-represents $Q_k$.

\begin{proof}[Proof of \Cref{lem:poly-lower-bound}]
    Let $C > 0$ be the constant from \Cref{lem:strict-sign-lb} and let $c > 0$ be a constant to be set later in the proof.
    Set $\eta = e^{-D/C}$ as in the statement.
    Suppose, seeking a contradiction, that there is a polynomial $p(s) = \sum_{|\alpha| \leq D} \lambda_\alpha s^\alpha$ such that $\max_\alpha |\lambda_\alpha| > e^{cD^2 + D\ln(40k)}$ and for all $s \in I_\eta^k$ we have $Q_k(s)p(s) \geq -1$.
    We construct a polynomial of degree at most $D$ that sign-represents $Q_k$ on $I_{2\eta}^k$, contradicting \Cref{lem:strict-sign-lb}.

    Define a $k$-variate polynomial $q$ by
    \[
        q(s) = \int_{[1,2]^k} p(\ldots, s_i x_i, \ldots) \,\dd x.
    \]
    If $s \in ([-1/2, -\eta] \cup [\eta, 1/2])^k$, then for all $x \in [1,2]^k$, $(\ldots, s_i x_i, \ldots) \in I_\eta^k$.
    Furthermore, for all $x \in [1,2]^k$ and $s$, we have $Q_k(\ldots, s_i x_i, \ldots) = Q_k(s)$.
    Thus by our assumption, for all $s \in ([-1/2, -\eta] \cup [\eta, 1/2])^k$ and $x \in [1,2]^k$, we have $Q_k(s) p(\ldots, s_i x_i, \ldots) \geq -1$ which implies
    \begin{equation} \label{eq:q-p-abs}
        Q_k(s) p(\ldots, s_i x_i, \ldots) \geq |p(\ldots, s_i x_i, \ldots)| - 2
    \end{equation}
    
    The coefficients of $p(\ldots, s_i x_i, \ldots)$ when viewed as a polynomial in $x$ are $\lambda'_\alpha = \lambda_\alpha s^\alpha$ for $|\alpha| \leq D$.
    For $s \in ([-1/2, -\eta] \cup [\eta, 1/2])^k$,
    \[
        \max_\alpha |\lambda'_\alpha| \geq \max_\alpha \left( |\lambda_\alpha| \cdot \eta^{|\alpha|} \right) > e^{cD^2 + D \ln(40k)} \eta^D.
    \]
    Integrating both sides of \eqnref{eq:q-p-abs} and applying \Cref{lem:coeff-int-ub}, we get
    \begin{align*}
        Q_k(s)q(s) &\geq \int_{[1,2]^k} |p(\ldots, s_i x_i, \ldots)| \,\dd x - 2 \\
        &\geq (40 k)^{-D} \max_\alpha |\lambda'_\alpha|  - 2 \\
        &> (40 k)^{-D} \cdot e^{cD^2 + D\ln(40k)} \cdot \eta^D - 2 \\
        &= e^{cD^2 - D^2/C} - 2.
    \end{align*}
    Setting $c$ to be a sufficiently large constant ensures that this quantity is positive.
    It follows that $q(s/2)$ is a degree-$D$ polynomial that sign-represents $Q_k$ on $I_{2\eta}^k$, which completes the proof.
\end{proof}

\section{Average Margin Bounds for \texorpdfstring{$\CHS \land \CHS$}{CHS and CHS}} \label{sec:continuous-bound}

We now combine \Cref{cor:sym-taylor} and \Cref{lem:poly-lower-bound} to obtain average margin upper bounds for the continuous problem $\CHS \land \CHS$.
More specifically, we will show that if $f$ and $g$ are normalized embeddings such that $B_{f,g}(x,y) := \langle f(x), g(y) \rangle$ sign-represents $\CHS \land \CHS$ on all inputs outside of a small band around the coordinate axes, then $B_{f,g}(x, y)$ must have small magnitude in expectation.
In subsequent sections we will ``round'' this to a lower bound for the discrete problem.

\begin{theorem} \label{thm:relaxation-lower-bound}
    Let $k \geq 2$ be an integer constant.
    There are constants $c, C > 0$ that depend on $k$ such that the following holds for all sufficiently large $d$. Let $f, g : (\mathbb{S}^{d-1})^k \to \R^r$ be measurable with $\|f(x)\|_2, \|g(x)\|_2 \leq 1$ for all $x$.
    Suppose that for all $x, y \in (\mathbb{S}^{d-1})^k$ with $\min_{i \in [k]} |\langle x_i, y_i \rangle| > e^{-Cd}$,
    \begin{equation} \label{eq:quad-corr}
        Q_k(\ldots, \langle x_i, y_i \rangle, \ldots) \cdot \langle f(x), g(y) \rangle \geq 0.
    \end{equation}
    Then, for independent, uniformly random $X, Y \in (\mathbb{S}^{d-1})^k$,
    \[
        \E \left[ |\langle f(X), g(Y) \rangle| \right] \leq e^{-cd}.
    \]
\end{theorem}

We give an informal overview of the proof of \Cref{thm:relaxation-lower-bound} before presenting it formally.
Let $X, Y$ be independent and uniformly random points in $(\mathbb{S}^{d-1})^k$, and let $S_i = \langle X_i, Y_i \rangle$.
Most of the technical work in the proof is to show an upper bound on the expectation of $|H(S)|$ where $H = H_{f,g}$ is the spherical average defined in \Cref{sec:spherical-avg}.
A simple calculation using concentration bounds on $|S_i|$ shows that this upper bound implies the desired upper bound on the expectation of $|\langle f(X), g(Y) \rangle|$.

To obtain the upper bound on $\mathbb{E}[|H(S)|]$ we look at the Taylor polynomial for $H$ of two different degrees.
Let $T_D(s) = \sum_{|\alpha| \leq D} \lambda_\alpha s^\alpha$ be the Taylor polynomial for $H$ with degree $D \approx d$.
The error bound in \Cref{cor:sym-taylor} shows that $T_D$ approximates $H$ well pointwise.
Averaging inequality~\eqnref{eq:quad-corr} over the sphere, we get $Q_k(s)H(s) \geq 0$ for all $s \in I_\eta^k$ for some $\eta = e^{-\Theta(D)}$.
Combining this with the error bound, we see that for all $s \in I_\eta^k$ we have $Q_k(s) T_D(e^{-K D} s) \geq -e^{-K'D^2}$ for some constants $K, K' > 0$.
We can thus apply \Cref{lem:poly-lower-bound} to a normalized version of $T_D(e^{-K D} s)$ to get an upper bound on every $|\lambda_\alpha|$.

The coefficient upper bound we obtain is roughly $|\lambda_\alpha| \leq e^{O(D^2) - KD(D - |\alpha|)}$.
Letting $K$ be sufficiently large, we see that this bound is exponentially small when $|\alpha|$ is small relative to $D$.
That is, for all $\alpha$ with $|\alpha| \leq D/2$, we have $|\lambda_\alpha| = e^{-\Omega(D^2)}$.
But that means that the Taylor polynomial of degree $D/2$ has small coefficients, and hence has low magnitude!
Using the error bounds from \Cref{cor:sym-taylor}, this time on the degree-$(D/2)$ Taylor polynomial, we get that $H$ itself must have low magnitude.
The upper bound on the expectation follows.

\begin{proof}[Proof of \Cref{thm:relaxation-lower-bound}]
    Let $f$ and $g$ be as in the theorem statement.
    Define $B(x,y) = \langle f(x), g(y) \rangle$ and let $H = H_{f,g}$ be as in \Cref{sec:spherical-avg}.
    Let $c', C' > 0$ be the constants from \Cref{lem:poly-lower-bound}.
    Set $D = \lfloor d/4 \rfloor$ and $\eta = e^{-D/C'}$.
    Let $K$ be a large constant to be determined later, and set $\rho = e^{-KD}$.
    The constant $K$ will depend on $c'$ and $k$.
    Note that if $d$ is sufficiently large then $\rho \leq 1/4$.
    Since $\log(\frac{1}{\rho\eta}) = O_{k, K, C'}(D) = O_{k, K, C'}(d)$, we can choose $C$ in the theorem statement to be a large constant (that depends on $K$) so that $e^{-Cd} \leq \rho\eta/2$.

    For every $s \in I_\eta^k$ we have $|\rho s_i| > e^{-Cd}$ for all $i \in [k]$.
    Thus inequality \eqnref{eq:quad-corr} implies that for all $s \in I_\eta^k$ and $x, y \in (\mathbb{S}^{d-1})^k$ with $\langle x_i, y_i \rangle = \rho s_i$, we have $Q_k(\rho s) B(x,y) \geq 0$.
    Averaging over $x$ and $y$ according to the experiment in the definition of $H$ and using the fact that $Q_k(\rho s) = Q_k(s)$, we get that $Q_k(s) H(\rho s) \geq 0$ for all $s \in I_\eta^k$.

    Let $T_D(s) = \sum_{|\alpha| \leq D} \lambda_\alpha s^\alpha$ be the degree-$D$ Taylor polynomial for $H$ at the origin.
    Since $\rho \leq 1/4$, for every $s \in I_\eta^k$ we have $\rho s \in [-1/4,1/4]^k$ and so using the error bound in \Cref{cor:sym-taylor},
    \begin{align*}
        Q_k(s) T_D(\rho s) &= Q_k(s) H(\rho s) + Q_k(s) (T_D(\rho s) - H(\rho s)) \\
        &\geq 0 - |T_D(\rho s) - H(\rho s)| \\
        &\geq - (4 \rho \|s\|_1)^{D+1} \geq -(4k\rho)^{D+1}.
    \end{align*}
    Applying \Cref{lem:poly-lower-bound} to the degree-$D$ polynomial
    \[
        p(s) = \frac{1}{(4k\rho)^{D+1}} \sum_{|\alpha| \leq D} \lambda_\alpha \rho^{|\alpha|} s^\alpha,
    \]
    we get that for all $\alpha$,
    \[
        |\lambda_\alpha| \leq e^{c' D^2 + D\ln(40k)} \cdot (4k)^{D+1} \cdot \rho^{D+1 - |\alpha|}.
    \]
    The above bound shows that if $|\alpha| \leq D/2$ then $|\lambda_\alpha|$ is exponentially small.
    That is, the coefficients of the \emph{degree-$(D/2)$} Taylor polynomial are small!
    We use this observation along with error bounds for that lower-degree Taylor polynomial to get an upper bound on the expectation of $|H|$ over random inner products.
    
    Let $m = \lfloor D/2 \rfloor$. Then we can set $K$ to be a large enough constant so that for all $|\alpha| \leq m$,
    \[
        |\lambda_\alpha| \leq \exp\left( c' D^2 + D\ln(40k) + (D+1)\ln(4k) -KD(\lceil D/2 \rceil + 1) \right) = e^{-\Omega(D^2)}.
    \]
    
    Notice that $T_m(s) := \sum_{|\alpha| \leq m} \lambda_\alpha s^\alpha$ is exactly the degree-$m$ Taylor polynomial for $H$ at the origin.
    Since $T_m$ has fewer than $(m+1)^k \leq D^k$ terms, for all $s \in [-1,1]^k$ and large $D$,
    \[
        |T_m(s)| \leq D^k e^{-\Omega(D^2)} = e^{-\Omega(D^2)}.
    \]
    So, applying \Cref{cor:sym-taylor} to $T_m$, for every $s \in [-1/(8k), 1/(8k)]^k$,
    \begin{align*}
        |H(s)| \leq e^{-\Omega(D^2)} + 2^{-(m+1)} = e^{-\Omega(D)} = e^{-\Omega(d)}.
    \end{align*}
    For all other $s \in [-1,1]^k$, we still have the weaker bound $|H(s)| \leq 1$.
    
    Let $X, Y$ be independent, uniformly random variables over $(\mathbb{S}^{d-1})^k$.
    For $i \in [k]$, let $S_i = \langle X_i, Y_i \rangle$.
    A union bound and the tail estimate~\eqnref{eq:score-upper-tail} yield
    \begin{align*}
        \mathbb{E}[|H(S)|] &\leq e^{-\Omega(d)} + \sum_{i \in [k]} \Pr[|S_i| > 1/(8k)] \cdot 1 \\
        &\leq e^{-\Omega(d)} + 8k^2 \left( 1 + \frac{1}{(8k)^2} \right)^{-(d-1)/2} = e^{-\Omega(d)}.
    \end{align*}

    To complete the proof, we show that the tail bounds in \Cref{lem:score-conc-bounds} imply that upper bounds on $\mathbb{E}[|H(S)|]$ yield upper bounds on $\mathbb{E}[|B(X,Y)|]$.
    Let $\mathsf{Bad}$ be the event that $\min_{i \in [k]} |S_i| \leq e^{-Cd}$.
    If the event does not occur, then the sign condition~\eqnref{eq:quad-corr} implies that $|B(X,Y)| = Q_k(S) B(X,Y)$.
    If the event does occur, then since $Q_k$ and $B$ are both bounded between $-1$ and $1$, $\big||B(X,Y)| - Q_k(S)B(X,Y)\big| \leq 2$ and thus $|B(X,Y)| \leq Q_k(S)B(X,Y) + 2$.
    Thus
    \begin{align*}
        \mathbb{E}[|B(X,Y)|] &\leq \mathbb{E}\big[Q_k(S)B(X,Y) (1 - \mathbbm{1}_\mathsf{Bad}) + ( Q_k(S)B(X,Y) + 2 )\mathbbm{1}_\mathsf{Bad}\big] \\
        &= \mathbb{E}\big[Q_k(S)B(X, Y) \big] + 2\Pr[\mathsf{Bad}].
        \numberthis \label{eq:B-expect-ub}
    \end{align*}
    By the tail bound~\eqnref{eq:score-lower-tail}, $\Pr[\mathsf{Bad}] \leq 2k \cdot e^{-Cd} \cdot \sqrt{d} = e^{-\Omega(d)}$.
    Also,
    \begin{align*}
        \mathbb{E}[Q_k(S) B(X, Y)] &= \E_{S} \big[ Q_k(S) \mathbb{E}[B(X,Y) | S] \big] \\
        &= \E_{S} [ Q_k(S) H(S) ] \\
        &\leq \mathbb{E}[|H(S)|] = e^{-\Omega(d)}.
    \end{align*}
    Plugging both bounds back into~\eqnref{eq:B-expect-ub}, $\mathbb{E}[|B(X, Y)|] = e^{-\Omega(d)}$, the desired upper bound.
\end{proof}

\section{Rounding to the Discrete Setting} \label{sec:rounding}

In this section, we show that the bound for $\CHS \land \CHS$ from \Cref{thm:relaxation-lower-bound} is robust enough to imply sign rank lower bounds for the discrete problem.
We state this transfer more generally for all discrete problems obtained from a continuous problem via sufficiently ``well-behaved'' rounding maps.
Our proof uses a generalization of Forster's theorem.

\begin{definition}[Average margin] \label{def:avg-margin}
    Let $A$ be a sign matrix with rows and columns indexed by $\mathcal{A}$ and $\mathcal{B}$ respectively.
    An $\emph{$r$-dimensional linear arrangement realizing}$ $A$ is a pair of collections of unit vectors $(\{u_a \in \mathbb{S}^{r-1}\}_{a \in \mathcal{A}}, \{v_b \in \mathbb{S}^{r-1}\}_{b \in \mathcal{B}})$ such that for all $a, b$, we have $A_{a,b} \cdot \langle u_a, v_b \rangle > 0$.
    Let $\mu$ be a distribution on $\mathcal{B}$.
    The \emph{average margin of $A$ with respect to $\mu$} is defined as
    \[
        m^{\mathrm{avg}}_\mu(A) = \sup_{u, v} \min_{a \in \mathcal{A}} \E_{b \sim \mu}[|\langle u_a, v_b \rangle|],
    \]
    where the supremum is over linear arrangements $(u, v)$ of arbitrary dimension that realize $A$.
\end{definition}

\begin{theorem}[Forster \cite{Forster02}, see also \cite{HatamiHPTZ22}] \label{thm:forster}
    For all sign matrices $A$ with rows and columns indexed by $\mathcal{A}$ and $\mathcal{B}$ respectively and distributions $\mu$ on $\mathcal{B}$,
    \[
        \sr(A) \geq \frac{1}{m^{\mathrm{avg}}_\mu(A)}.
    \]
\end{theorem}

\begin{theorem} \label{thm:rounding}
    Let $\mathcal{X}$ and $\mathcal{Y}$ be probability spaces, let $X$ and $Y$ be independent samples from $\mathcal{X}$ and $\mathcal{Y}$ respectively, and let $E \subseteq \mathcal{X} \times \mathcal{Y}$ be a measurable set.

    Let $A$ be a sign matrix with rows and columns indexed by $\mathcal{A}$ and $\mathcal{B}$. Let $\Pi_1 : \mathcal{X} \to \mathcal{A}$ and $\Pi_2 : \mathcal{Y} \to \mathcal{B}$ be measurable functions. Then
    \begin{equation} \label{eq:sr-sup}
        \sr(A) \ge \frac{1}{\sup_{f, g} \E_{X,Y} [ | \langle f(X), g(Y) \rangle | ]}
    \end{equation}
    where the supremum is taken over all finite dimensions $r$, measurable arrangements $f : \mathcal{X} \to \R^r$ and $g : \mathcal{Y} \to \R^r$  satisfying $\|f(x)\|_2, \|g(y)\|_2 \leq 1$ for all $x \in \mathcal{X}, y\in\mathcal{Y}$ and 
    \[A_{\Pi_1(x), \Pi_2(y)} \cdot \langle f(x), g(y) \rangle \geq 0\]
    whenever $(x, y) \notin E$.
\end{theorem}
\begin{proof}
    Let $X, Y$ be independent samples from $\mathcal{X}, \mathcal{Y}$.
    Define the probability distribution $\mu$ on $\mathcal{B}$ by the mass function $\mu(b) = \Pr[\Pi_2(Y) = b]$.
    Let $(u,v)$ be an $r$-dimensional linear arrangement that realizes $A$ as in \Cref{def:avg-margin}.
    Define $f : \mathcal{X} \to \R^r$ and $g : \mathcal{Y} \to \R^r$ by $f(x) = u_{\Pi_1(x)}$ and $g(y) = v_{\Pi_2(y)}$.
    Since each $u_a$ is a unit vector, $\|f(x)\|_2 \leq 1$ for all $x \in \mathcal{X}$.
    Similarly, $\|g(y)\|_2 \leq 1$ for all $y \in \mathcal{Y}$.
    Also, since $(u,v)$ realizes $A$, for every $x, y$ we have
    \[
        A_{\Pi_1(x), \Pi_2(y)} \cdot \langle f(x), g(y) \rangle = A_{\Pi_1(x), \Pi_2(y)} \cdot \langle u_{\Pi_1(x)}, v_{\Pi_2(y)} \rangle \geq 0.
    \]
    By the definitions of $f$, $g$, and $\mu$,
    \[
        \E_{X,Y} [ | \langle f(X), g(Y) \rangle | ] = \E_{X,Y} [ | \langle u_{\Pi_1(X)}, v_{\Pi_2(Y)} \rangle | ] = \E_{b \sim \mu, X} [ | \langle u_{\Pi_1(X)}, v_b \rangle | ] \geq \min_{a \in \mathcal{A}} \E_{b \sim \mu} [ | \langle u_{a}, v_b \rangle | ].
    \]
    Thus $\min_{a \in \mathcal{A}} \E_{b \sim \mu} [ | \langle u_{a}, v_b \rangle | ] \leq \sup_{f,g} \E_{X,Y} [ | \langle f(X), g(Y) \rangle | ]$ where the supremum is over $f$ and $g$ that satisfy the conditions in the theorem statement.
    
    Since $(u, v)$ was an arbitrary linear arrangement realizing $A$, we get 
    \[
        m_\mu^\mathrm{avg}(A) = \sup_{u,v} \min_{a \in \mathcal{A}} \E_{b \sim \mu} [|\langle u_a, v_b \rangle|] \leq \sup_{f,g} \E_{X,Y} [ | \langle f(X), g(Y) \rangle | ].
    \]
    The theorem statement follows by \Cref{thm:forster}.
\end{proof}

Our main results apply \Cref{thm:rounding} with the following setting of parameters.
\begin{corollary} \label{cor:intersection-rounding}
    There are constants $c, C > 0$ such that the following holds for sufficiently large $d$.
    Let $B$ be a sign matrix with rows and columns indexed by $\mathcal{A}$ and $\mathcal{B}$.
    Suppose there are measurable maps $\pi_1 : \mathbb{S}^{d-1} \to \mathcal{A}$ and $\pi_2 : \mathbb{S}^{d-1} \to \mathcal{B}$ such that for all $x, y \in \mathbb{S}^{d-1}$ with $|\langle x, y \rangle| > e^{-Cd}$ we have $B_{\pi_1(x), \pi_2(y)} = \sgn \langle x, y \rangle$.
    Then $\sr(B \land B) \geq e^{cd}$.
\end{corollary}
\begin{proof}
    Let $c, C > 0$ be the constants from \Cref{thm:relaxation-lower-bound}.
    We apply \Cref{thm:rounding} with the following parameters.
    \begin{enumerate}
        \item $\mathcal{X} = \mathcal{Y} = (\mathbb{S}^{d-1})^2$ equipped with the uniform measure.
        \item $E = \big\{ (x, y) \in (\mathbb{S}^{d-1})^2 \times (\mathbb{S}^{d-1})^2 \mid \min_{i \in \{1,2\}} \{ |\langle x_i, y_i \rangle| \} \leq e^{-Cd} \big\}$.
        \item $A = B \land B$ indexed by $\mathcal{A}^2$ and $\mathcal{B}^2$.
        \item $\Pi_1(x) = (\pi_1(x_1), \pi_1(x_2))$ and $\Pi_2(y) = (\pi_2(y_1), \pi_2(y_2))$.
    \end{enumerate}
    
    Let $f : \mathcal{X} \to \R^r$, $g : \mathcal{Y} \to \R^r$ be an arrangement such that for all $x \in \mathcal{X}$ and $y \in \mathcal{Y}$, we have $\|f(x)\|_2, \|g(y)\|_2 \leq 1$ and furthermore if $(x,y) \notin E$ then $A_{\Pi_1(x), \Pi_2(y)} \cdot \langle f(x), g(y) \rangle \geq 0$.
    We have $A_{\Pi_1(x), \Pi_2(y)} = B_{\pi_1(x_1), \pi_2(y_1)} \land B_{\pi_1(x_2), \pi_2(y_2)}$.
    By our hypothesis, this is equal to $\sgn \langle x_1, y_1 \rangle \land \sgn \langle x_2, y_2 \rangle = Q_2(\langle x_1, y_1 \rangle, \langle x_2, y_2 \rangle)$.
    Thus by \Cref{thm:relaxation-lower-bound}, if $d$ is sufficiently large then $\mathbb{E}[|\langle f(X), g(Y) \rangle|] \leq e^{-cd}$ where $X, Y$ are independent uniformly random variables over $(\mathbb{S}^{d-1})^2$.
    It follows that the supremum in inequality~\eqref{eq:sr-sup} is bounded above by $e^{-cd}$.
    Hence by \Cref{thm:rounding}, $\sr(B \land B) \geq e^{cd}$.
\end{proof}

\section{Proof of the Main Theorems} \label{sec:main}

\subsection{The Sign Rank of \texorpdfstring{$\HS \land \HS$}{HS and HS}}

\begin{theorem} \label{thm:HS-lower-bound}
    There exist universal constants $c, C > 0$ such that if $M \geq \lceil e^{Cd} \rceil$ then
    $\sr(\HS_{M,d}) \leq d+1$ and for sufficiently large $d$, $\sr(\HS_{M,d} \land \HS_{M,d}) \geq e^{cd}$.
    Consequently, if $n$ is the number of bits required to encode each row and column of $\HS_{M,d}$ for $M = \lceil e^{Cd} \rceil$, then $U(\HS_{M,d}) = O(\log n)$ whereas $U(\HS_{M,d} \land \HS_{M,d}) = \Omega(\sqrt{n})$.
\end{theorem}
\begin{proof}
    Let $C$ be a constant to be set later, and set $M = \lceil e^{Cd} \rceil$.
    The matrix $\HS_{M,d}$ agrees entry-wise in sign to the matrix $A_{a,b} = \langle a, b \rangle - 1/2$. Since the inner product has rank $d$ and rank is subadditive, we have $\rank(A) \leq d + 1$ and hence $\sr(\HS_{M,d}) \leq d + 1$.

    Let $\pi : \mathbb{S}^{d-1} \to \{-M, \ldots, M\}^d$ be $\pi(x_1, \ldots, x_d) = (\mathrm{round}(Mx_1), \ldots, \mathrm{round}(Mx_d))$ where $\mathrm{round}(\cdot)$ evaluates to the integer closest to the input.
    Let $c, C' > 0$ be the constants from \Cref{cor:intersection-rounding}.
    We show that when $d$ is sufficiently large, for all $x, y \in \mathbb{S}^{d-1}$ such that $|\langle x, y \rangle| > e^{-C'd}$, $(\HS_{M,d})_{\pi(x), \pi(y)} = \sgn \langle x, y \rangle$.
    \Cref{cor:intersection-rounding} then implies that $\sr(\HS_{M,d} \land \HS_{M,d}) \geq e^{cd}$ for large $d$.
    
    Notice that $(\HS_{M,d})_{\pi(x), \pi(y)} = \sgn\big( \langle \pi(x), \pi(y) \rangle - \frac{1}{2} \big) = \sgn\big( \frac{1}{M^2} (\langle \pi(x), \pi(y) \rangle - \frac{1}{2}) \big)$.
    Thus it suffices to show that $\left| \frac{1}{M^2} \big( \langle \pi(x), \pi(y) \rangle - \frac{1}{2} \big) - \langle x, y \rangle \right| < e^{-C'd}$.
    For $z \in \mathbb{S}^{d-1}$ define $e_z = \frac{\pi(z)}{M} - z$ so that $\pi(z)/M = z + e_z$. We have
    \[
        \frac{1}{M^2} \left( \langle \pi(x), \pi(y) \rangle - \frac{1}{2} \right) = \langle x + e_x, y + e_y \rangle - \frac{1}{2M^2}.
    \]
    Expanding the inner product, rearranging, and using the triangle inequality gives us
    \begin{align*}
        \left| \frac{1}{M^2} \left( \langle \pi(x), \pi(y) \rangle - \frac{1}{2} \right) - \langle x, y \rangle \right|
        &\leq |\langle x, e_y \rangle| + |\langle e_x, y \rangle| + |\langle e_x, e_y \rangle| + \frac{1}{2M^2} \\
        &\leq \|e_y\|_2 + \|e_x\|_2 + \|e_x\|_2 \|e_y\|_2 + \frac{1}{2M^2}.
    \end{align*}
    The second line holds due to the Cauchy--Schwarz inequality and the fact that $\|x\|_2, \|y\|_2 \leq 1$.
    Now note that for all $z$ and $i \in [d]$, $|\pi(z)_i - Mz_i| \leq 1/2$. It follows that $\|e_z\|_2 \leq \sqrt{d}/(2M)$.
    Thus
    \[
        \left| \frac{1}{M^2} \left( \langle \pi(x), \pi(y) \rangle - \frac{1}{2} \right) - \langle x, y \rangle \right| \leq \frac{\sqrt{d}}{M} + \frac{d}{4M^2} + \frac{1}{2M^2}.
    \]
    Setting $C$ (and hence $M \geq \lceil e^{Cd} \rceil$) to be large enough, we get that for sufficiently large $d$ the right hand side is less than $e^{-C'd}$, proving the required inequality.
    
    The number of bits required to index a row or column of $\HS_{M,d}$ is $n = O(d \log M) = O(d^2)$.
    By \Cref{thm: ps}, $U(\HS_{M,d}) = O(\log d) = O(\log n)$ and $U(\HS_{M,d} \land \HS_{M,d}) = \Omega(\sqrt{n})$.
\end{proof}

\subsection{Matrices of Random Unit Vectors}

\begin{theorem} \label{thm:gaussian-lb}
    There exist constants $c, C > 0$ such that the following holds.
    Let $N = \lceil e^{Cd^2} \rceil$ and let $u_1, \ldots, u_N, v_1, \ldots, v_N$ be independently sampled uniformly random points on $\mathbb{S}^{d-1}$. Let $G_{i,j} = \sgn \langle u_i, v_j \rangle$.
    Then with probability $1 - o(1)$, $\sr(G) \leq d$ whereas $\sr(G \land G) \geq e^{cd}$.
\end{theorem}
\begin{proof}
    With probability one, $u_i, v_j \neq \vec{0}$ and $\langle u_i, v_j \rangle \neq 0$ for all $i,j$.
    We condition on this event for the rest of the proof.
    Note that the matrix $G$ is sign-represented by the matrix $G'_{i,j} = \langle u_i, v_j \rangle$. Since the inner product has rank $d$, we get $\rank(G') = d$ and thus $\sr(G) \leq d$.

    Let $c', C' > 0$ be the constants from \Cref{cor:intersection-rounding}, and set $\delta = \frac{1}{4}e^{-C'd}$.
    As long as the constant $C$ in the definition of $N$ is large enough, by \Cref{lem:delta-net} and the union bound, with probability $1 - o(1)$ every point of $\mathbb{S}^{d-1}$ is within Euclidean distance $\delta$ of some $u_i$ and also some $v_j$.
    Condition on this event.
    Let $\pi_1, \pi_2 : \mathbb{S}^{d-1} \to [N]$ be the maps that send $x$ to the $i$ that minimizes $\|x - u_i\|_2$ and $\|x - v_i\|_2$ respectively (if there is a tie simply pick the lower index).

    Let $x, y \in \mathbb{S}^{d-1}$ be arbitrary, and set $i = \pi_1(x)$ and $j = \pi_2(y)$.
    By the triangle inequality and Cauchy--Schwarz,
    \begin{align*}
        |\langle u_i, v_j \rangle - \langle x, y \rangle|
        &\leq |\langle u_i - x, v_j \rangle| + |\langle x, v_j - y \rangle| \\
        &\leq \| u_i - x \|_2 + \| v_j - y \|_2 \\
        &\leq 2\delta < e^{-C'd}.
    \end{align*}
    Thus if $|\langle x, y \rangle| > e^{-C'd}$ then $\langle u_i, v_j \rangle$ has the same sign as $\langle x, y \rangle$.
    The sign rank lower bound follows by \Cref{cor:intersection-rounding}.
\end{proof}

\section{On the Tightness of Our Bounds} \label{sec:tightness}

In this section, we show that our average margin upper bound in continuous space (\Cref{thm:relaxation-lower-bound}) and our main sign rank lower bounds in the discrete setting are essentially tight for their respective problems.

\begin{theorem}[Optimality of~\Cref{thm:relaxation-lower-bound}] \label{thm:cont-opt-weak}
For all $k\ge 1$ and sufficiently large $d$, there are measurable functions $f, g : (\mathbb{S}^{d-1})^k \to [0, 1]$ such that
\[Q_k(\dots, \langle x_i, y_i\rangle, \dots) \cdot f(x) \cdot g(y) \ge 0\]
for all $x, y \in (\mathbb{S}^{d-1})^k$ and
\[\mathbb{E}[|f(X) g(Y)|] = 2^{-\Theta(d)}\]
for independent, uniformly random $X, Y \sim (\mathbb{S}^{d-1})^k$.
\end{theorem}

That is, even for $(k=2)$-fold intersections, embedding dimension $r = 1$, and the requirement of weak sign representation on all of $((\mathbb{S}^{d-1})^k)^2$, there are arrangements $f, g$ for which the bound of~\Cref{thm:relaxation-lower-bound} is optimal.

\begin{proof}
    Consider the spherical cap $B = \{z \in \mathbb{S}^{d-1} \mid z_1 \ge \sqrt{3}/2\}$. Observe that if $x, y \in B$, then $\langle x, y \rangle \ge \frac{3}{4} - \frac{1}{4} = \frac{1}{2}$ by applying Cauchy--Schwarz to the remaining coordinates. Set
    \[f(x_1, \dots, x_k) = \begin{cases}
        1 & \text{if } x_1 \in B\\
        0 & \text{otherwise,}
    \end{cases}
    \qquad
    g(y_1, \dots, y_k) = \begin{cases}
        1 & \text{if } y_1 \in B \\
        0 & \text{otherwise.}
    \end{cases}\]
    Then $f(x) g(y)$ is nonzero only if $\langle x_1, y_1\rangle > 0$, so weak sign agreement holds. Meanwhile,
    if we let $\mu$ denote the uniform spherical measure, $\mathbb{E}[|f(X)g(Y)|] = (\mu(B))^2$ by independence. \Cref{lem:score-conc-bounds} shows that $\mu(B) = 2^{-\Theta(d)}$.
\end{proof}

A limitation of~\Cref{thm:cont-opt-weak} is that the arrangement $f, g$ constructed only weakly sign-represents $Q_k$. The fact that~\Cref{thm:relaxation-lower-bound} implies a small average margin even for weak sign representations is not really needed in its downstream applications. Nevertheless, we can show that it is essentially tight even for arrangements that are strong sign representations.

\begin{theorem}[Optimality of~\Cref{thm:relaxation-lower-bound} for strong sign representations] \label{thm:cont-opt-strong}
For all $\varepsilon > 0$, $k\ge 2$ and sufficiently large $d$, there exists a dimension $r$ and measurable functions $f, g : (\mathbb{S}^{d-1})^k \to \mathbb{R}^r$ such that $\|f(x)\|_2, \|g(x)\|_2 \le 1$ for every $x$ and
\[Q_k(\dots, \langle x_i, y_i\rangle, \dots) \cdot \langle f(x) , g(y)\rangle > 0\]
for all $x, y \in (\mathbb{S}^{d-1})^k$ with $\min_{i \in [k]} |\langle x_i, y_i\rangle| > \varepsilon$, and such that 
\[\mathbb{E}[|\langle f(X), g(Y)\rangle |] = 2^{-\Theta(d)}\]
for independent, uniformly random $X, Y \sim (\mathbb{S}^{d-1})^k$.
\end{theorem}

\begin{proof}
    Let $f_0, g_0$ be the arrangement constructed in~\Cref{thm:cont-opt-weak} that weakly sign-represents $Q_k$ and has exponentially small average margin $m_0 = \mathbb{E}[|f_0(X)g_0(Y)|] = 2^{-\Theta(d)}$. Let $f_1, g_1$ be an arbitrary arrangement, in some sufficiently large finite dimension, that strongly sign-represents $Q_k$ whenever $\min_{i \in [k]} |\langle x_i, y_i\rangle| > \varepsilon$ and is normalized to $1$. (The fact that such an arrangement exists follows from the proof of Lemma~\ref{lem:sr-lb-generic} below.) For $\delta = m_0/4$, define
    \[f(x) = (\sqrt{1-\delta}f_0(x), \sqrt{\delta} f_1(x)), \qquad g(y) = (\sqrt{1-\delta}g_0(y), \sqrt{\delta} g_1(y)).\]
    The norm of each function is always at most $1$ and their inner product is
    \[\langle f(x), g(y)\rangle = (1-\delta) \langle f_0(x), g_0(y)\rangle + \delta \langle f_1(x), g_1(y)\rangle.\]
    So the fact that $f_0, g_0$ weakly sign-represent $Q_k$ everywhere and $f_1, g_1$ strongly sign-represent $Q_k$ whenever $\min_{i \in [k]} |\langle x_i, y_i\rangle| > \varepsilon$ means $f, g$ do the latter as well.

    Now by the triangle inequality, $\mathbb{E}[|\langle f(X), g(Y)\rangle|] \in (1-\delta)\mathbb{E}[|f_0(X) g_0(Y)|] \pm \delta\mathbb{E}[|\langle f_1(X), g_1(Y)\rangle|] \in (1-\delta) m_0 \pm \delta = 2^{-\Theta(d)}$.
\end{proof}

We now prove the optimality of our main sign rank lower bounds. Recall our notation $I_\eta = [-1, -\eta] \cup [\eta, 1]$.

\begin{lemma} \label{lem:sign-rep}
    For every $\eta \in (0, 1/2)$ and $k \geq 1$, there is a polynomial $q : I_\eta^k \to \mathbb{R}$ of degree $O(k^2\log(1/\eta))$ such that $Q_k(s) \cdot q(s) > 0$ for every $s \in I_\eta^k$.
\end{lemma}

\begin{proof}
    The proof proceeds by combining Newman's classic rational approximation to the sign function with the observation of Beigel, Reingold, and Spielman~\cite{BRS95} that rational approximants can be combined to sign-represent an intersection. In more detail, Sherstov's analysis~\cite[Theorem 2.4]{Sherstov13} of Newman's construction~\cite{newman1964approximation} shows that there are real polynomials $P, R$ of degree $D \leq \lceil 1 + 2k \log(1/\eta) \rceil$ such that
    \[\left| \frac{P(s)}{R(s)} - \sgn(s)\right| \le \frac{1}{2k}\]
    for all $s \in I_\eta$ and $R$ never vanishes on $[-1, 1]$. Now the function
    \[(k-1) + \frac{P(s_1)}{R(s_1)} + \frac{P(s_2)}{R(s_2)} + \dots + \frac{P(s_k)}{R(s_k)}\]
    sign-represents $Q_k(s_1, \dots, s_k) = \sgn s_1 \land \dots \land \sgn s_k$. Placing everything over a common denominator lets us write this as $p(s_1, \dots, s_k) / r(s_1, \dots, s_k)$ where the polynomials $p, r$ each have total degree at most $kD$. Since $p \cdot r$ always has the same sign as $p/r$, we may take $q = p\cdot r$ to be our final polynomial.
\end{proof}

\begin{lemma} \label{lem:sr-lb-generic}
    Let $\eta \in (0, 1/2)$, $k \geq 1$, and let $\mathcal{A}, \mathcal{B} \subseteq \mathbb{R}^d$ be finite sets such that $\eta \le |\langle a, b \rangle| \le 1$ for all $a \in \mathcal{A}$ and $b \in \mathcal{B}$. Let $A \in \{-1, 1\}^{\mathcal{A} \times \mathcal{B}}$ be the sign matrix defined by $A_{a, b} = \sgn(\langle a, b \rangle)$. Then $\sr(\land_{k} A) \le \binom{kd + D}{D}$ for $D = O(k^2 \log (1/\eta))$.
\end{lemma}

\begin{proof}
    Let $D = O(k^2 \log (1/\eta))$ and $q(s)$ be the polynomial of degree $D$ guaranteed by~\Cref{{lem:sign-rep}} to sign-represent $Q_k$. Then
    \[(\land_k A)_{(a_1, \dots, a_k), (b_1, \dots, b_k)} = Q_k(\sgn \langle a_1, b_1\rangle, \dots, \sgn \langle a_k, b_k\rangle) = \sgn q(\langle a_1, b_1 \rangle, \dots, \langle a_k, b_k\rangle) \]
    The right-hand quantity is the sign of a polynomial of total degree at most $D$ in Alice's variables $a_1, \dots, a_k$. So we may write that polynomial as
    \[q(\langle a_1, b_1 \rangle, \dots, \langle a_k, b_k\rangle) = \sum_{|\alpha| \le D} (a_1, \dots, a_k)^\alpha q_\alpha(b_1, \dots, b_k)\]
    for some polynomials $q_\alpha$. Each summand is a rank-one matrix, so their sum is a matrix of rank at most $\binom{kd + D-1}{D} + \binom{kd+D-2}{D-1} + \dots + 1 \le \binom{kd + D}{D}$.
\end{proof}

\begin{theorem}[Tightness of~\Cref{thm:HS-lower-bound}] \label{thm:HS-tight}
For all positive integers $d$ and $M$,
\[\sr(\HS_{M, d}) \ge d,\]
\[\sr(\HS_{M, d} \land \HS_{M, d}) \le 2^{O(d + \log M)}.\]
In particular, the lower bound of~\Cref{thm:HS-lower-bound} is tight when $M = 2^{\Theta(d)}$.
\end{theorem}

\begin{proof}
    For the first claim, let us restrict our attention to the submatrix consisting of rows in $\{-1, 1\}^d$ and columns $e_1, \dots, e_d$ representing the $d$ standard basis vectors. All $2^d$ patterns of signs appear in this submatrix, showing it has VC dimension $d$. Hence, the sign rank of this matrix is also at least $d$~\cite{PS86}.

    For the second claim, observe that for all $a, b \in \{-M, \dots, M\}^d$, their inner product $\langle a, b\rangle$ is an integer between $-dM^2$ and $dM^2$. Normalizing shows that $\HS_{M, d}$ is equivalent to a sign matrix meeting the conditions of~\Cref{lem:sr-lb-generic} in dimension $d+1$ (to account for the additive shift of $-1/2$ in its definition) and with $\eta = 1/(2dM^2 + 1)$. Taking $k = 2$ and $D = O(\log(2dM^2 + 1))$ shows that the sign rank of $\HS_{M, d} \land \HS_{M, d}$ is at most
    \[\binom{2d + D + 2}{D} \le 2^{O(d + \log M)}.\]
\end{proof}

\begin{theorem}[Tightness of~\Cref{thm:gaussian-lb}]
Let $u_1, \dots, u_N, v_1, \dots, v_N$ be independently sampled standard Gaussians on $\R^d$ and define the matrix $G$ by $G_{i, j} = \sgn\langle u_i, v_j\rangle$. Then with probability $1-o(1)$,
\[\sr(G \land G) \le 2^{O(d \log \log N)}.\]
In particular, the lower bound of~\Cref{thm:gaussian-lb} is tight up to a factor of $\log d$ in the exponent for $\lceil e^{Cd^2}\rceil \le N \le 2^{d^{O(1)}}$.
\end{theorem}

\begin{proof}
    Let $\hat{u_i} = u_i / \|u_i\|_2$ and $\hat{v_i} = v_i / \|v_i\|_2$ be the normalized sampled points, which are independent uniform points on the unit sphere $\mathbb{S}^{d-1}$. By~\Cref{lem:score-conc-bounds}, for every $\eta > 0$ we have
    \[\Pr[|\langle\hat{u_i}, \hat{v_j}\rangle| \le \eta] \le 2\eta\sqrt{d}\]
    for each pair $i, j$. By union bounding over the (at most) $N^2$ pairs of points, we get that
    \[\Pr\left[\min_{i, j} |\langle\hat{u_i}, \hat{v_j}\rangle| > \eta\right] \ge 1-\frac{2}{Nd} = 1-o(1)\]
    for $\eta = \frac{1}{2N^3\cdot d^{3/2}}$. If this lower bound on the absolute inner products holds, then $G$ is equivalent to a matrix meeting the conditions of~\Cref{lem:sr-lb-generic} with this value of $\eta$. Taking $k = 2$ and $D = O(\log N + \log d)$ shows that the sign rank of $G \land G$ is at most
    \[\binom{2d + D}{D} = \binom{2d + D}{2d} \le 2^{O(d\log \log N)}.\]
\end{proof}

\section{Applications} \label{sec:applications}

For simplicity in describing bitwise arithmetic, it is helpful in this section to regard Boolean functions as taking inputs in $\{0, 1\}^n$ (where $0$ is logical false and $1$ is logical true) to outputs in $\{-1, 1\}$ (where $1$ is logical false and $-1$ is logical true). Our workhorse communication problem is the following variant of the bounded integer halfspace problem defined on nonnegative inputs. Define the function $\widetilde{HS}_{M,d}(a, b)$ on inputs $a, b \in \{0, \dots, 2M\}^d$ by $\widetilde{HS}_{M,d}(a, b) = \HS_{M, d}(a - M \mathbf{1}, b-M \mathbf{1}) = \sgn(\langle a, b\rangle - M\langle a + b, \mathbf{1}\rangle + dM^2- 1/2)$. Observe that we also have for every $a, b \in \{-M, \dots, M\}^d$ that $\HS_{M, d}(a, b) = \widetilde{HS}_{M,d}(a + M\mathbf{1}, b + M\mathbf{1})$. So the problems are equivalent and in particular have the same sign rank:
\begin{align*}
    &\sr(\widetilde{HS}_{M,d}) = \sr(\HS_{M, d}) = O(d), \\
    &\sr(\widetilde{HS}_{M,d} \land \widetilde{HS}_{M,d}) = \sr(\HS_{M, d} \land \HS_{M, d}) = 2^{\Omega(d)}
\end{align*}
for appropriate $M = 2^{\Theta(d)}$, by Theorem~\ref{thm:HS-lower-bound}.

Our applications rely on the following lemma that lets us express $\widetilde{HS}_{M,d} $ as a halfspace problem over Boolean inputs.
In subsequent subsections we apply the lemma and our main theorem to obtain the corollaries from Section~\ref{sec:intro-apps}.
We will recall the statements of those corollaries for the reader's convenience.

\begin{lemma} \label{lem:binary-hs-rep}
  For natural numbers $M, d$ let $n = d \cdot \lceil \log(2M+1)\rceil$. There exists a set $S$ of at most $O(d \log^2 M)$ pairs of indices $(i, j) \in \{0, 1, \dots, n\}^2$ such that for every
  $a, b \in \{0, \dots, 2M\}^d$,
  \[\widetilde{HS}_{M,d}(a, b) = \sgn\left(\sum_{(i, j)\in S} c_{(i, j)} x_i y_j\right)\]
  where $x = (x_1, \dots, x_n), y = (y_1, \dots, y_n)$ are the binary encodings of $a, b$ respectively, and $x_0 = y_0 = 1$.
\end{lemma}

\begin{proof}
    Let $a = (a_1, \dots, a_d) \in \{0, \dots, 2M\}^d$ and $b = (b_1, \dots, b_d) \in \{0, \dots, 2M\}^d$. Write each component, e.g., $a_i$ in its standard representation as an $m$-bit number, where $m = \lceil \log(2M+1)\rceil$, so $a_i^{(m-1)}a_i^{(m-2)}\dots a_i^{(0)}$ represents the integer $a_i = \sum_{j=0}^{m-1} 2^j a_i^{(j)}$. Then
        \begin{align*}
        \widetilde{HS}_{M,d}(a, b) &= \sgn\left(\sum_{i = 1}^d a_i b_i - M \sum_{i=1}^d(a_i + b_i) + dM^2 - \frac{1}{2}\right) \\
        &= \sgn\left(\sum_{i = 1}^d \left(\sum_{j=0}^{m-1} 2^j a_i^{(j)} \right) \left(\sum_{j=0}^{m-1} 2^j b_i^{(j)} \right) - M \sum_{i=1}^d \sum_{j=0}^{m-1} 2^j (a_i^{(j)} + b_i^{(j)}) + dM^2 - \frac{1}{2}  \right) \\
        &= \sgn\left(\sum_{i=1}^d \sum_{0 \le j, \ell \le m-1} 2^{j + \ell} a_i^{(j)} b_i^{(\ell)}  - M \sum_{i=1}^d \sum_{j=0}^{m-1} 2^j (a_i^{(j)} + b_i^{(j)}) + dM^2 - \frac{1}{2}  \right).
    \end{align*}
\end{proof}

\subsection{Threshold-of-Majority Circuit Size}

\getkeytheorem{circuit-sep}

\begin{proof}
    We will take the function $F$ in the statement to be $\widetilde{HS}_{M,d}$ for $M = 2^{\Theta(d)}$. We begin by showing that this function is computed by small Threshold-of-Majority circuits; indeed, it suffices for the bottom majority gates to compute the AND function on at most $2$ bits.

    \begin{lemma}
        For every $M \ge 2, d$, the function $\widetilde{HS}_{M,d}$ is computed by a Threshold-of-AND$_2$ circuit of size $O(d\log^2 M)$.
    \end{lemma}
    
    \begin{proof}
        Consider the representation of $\widetilde{HS}_{M,d}$ given by~\Cref{lem:binary-hs-rep}. Each term $x_i y_j$ is a pairwise AND over the domain $\{0, 1\}$, so this representation shows that $\widetilde{HS}_{M,d}$ is computed by a threshold of $O(d \log^2M)$ ANDs, each of at most two variables.
    \end{proof}
    
    Meanwhile, the fact that $\sr(\widetilde{HS}_{M,d} \land \widetilde{HS}_{M,d}) \ge 2^{\Omega(d)}$, implies this intersection has Threshold-of-Majority circuit size $2^{\Omega(d)}$.
\end{proof}

\subsection{Dimension Complexity of Intersections of Halfspaces}

\getkeytheorem{halfspace-learning}
\begin{proof}
    Set $M = 2^{\Theta(d)}$ as in the statement of~\Cref{thm:HS-lower-bound}, in dimension $d/2$.
    
    The first statement holds by taking $\mathcal{C}_d  = \{f_{a_1, a_2} : (\{0, \dots, 2M\}^{d/2})^2 \to \{-1, 1\} \mid (a_1, a_2) \in (\{0, \dots, 2M\}^{d/2})^2 \}$ where each $f_{a_1, a_2}(b_1, b_2) =\widetilde{HS}_{M,d/2}(a_1, b_1)\land \widetilde{HS}_{M,d/2}(a_2, b_2)$.
    
    The second statement then follows from~\Cref{lem:binary-hs-rep}, which shows that each $f_{a_1, a_2}(b_1, b_2)$ is equivalent to an intersection of Boolean halfspaces defined over $n = \Theta(d^2)$ variables.
\end{proof}

\subsection{Approximate vs.\ Exact Sign Rank}

\getkeytheorem{approx-vs-exact}

\begin{proof}
    Recall that the $\varepsilon$-approximate sign rank of Boolean matrix $A \in \{-1, 1\}^{\mathcal{A} \times \mathcal{B}}$ is the least dimension $r$ in which there is a collection of vectors $\{v_b\}_{b \in \mathcal{B}} \subseteq \mathbb{R}^{r}$ such that for every $a \in \mathcal{A}$ and distribution $\nu$ over $\mathcal{B}$, there exists a vector $u_a \in \mathbb{R}^r$ such that $\Pr_{b \sim \nu}[\sgn \langle u_a, v_b\rangle \ne A_{a, b}] \le \varepsilon$. 
    
    Let $M = \lceil e^{Cd} \rceil$ and $\mathcal{A} = \mathcal{B} = (\{-M, \dots, M\}^d)^2$ where $C$ is the constant from \Cref{thm:HS-lower-bound}. Let $A$ be the communication matrix of $\HS_{M, d} \land \HS_{M, d}$. Explicitly, we have for $(a_1, a_2) \in \mathcal{A}$ and $(b_1, b_2) \in \mathcal{B}$, that
    \[A_{(a_1, a_2), (b_1, b_2)} = \left(\sgn \left(\langle a_1, b_1\rangle - \frac{1}{2}\right) \land \sgn \left(\langle a_2, b_2 \rangle - \frac{1}{2}\right)\right).\]
    
    We now show that this matrix $A$ has $(1/3)$-approximate sign rank at most $2d+1$, following a similar argument from~\cite{ChornomazMW25}. For each $(b_1, b_2) \in \mathcal{B}$, define $v_{(b_1, b_2)} = (b_1, b_2, 1) \in \mathbb{R}^{2d+1}$. Fix a distribution $\nu$ over $\mathcal{B}$ and $(a_1, a_2) \in \mathcal{A}$. For each $i \in \{1, 2\}$, let $E_i$ denote the event that $\HS_{M, d}(a_i, b_i) = A_{(a_1, a_2), (b_1, b_2)}$. Let $E_3$ be the event that $A_{(a_1, a_2), (b_1, b_2)} = 1$ (logical false). By the definition of $A$, we have that for every $b \in \mathcal{B}$ at least two out of three of the events $E_1, E_2, E_3$ hold. In particular, $\mathbb{E}_{b \sim \nu}[\mathbbm{1}[E_1] + \mathbbm{1}[E_2] + \mathbbm{1}[E_3]] \ge 2$, so $\Pr_{b \sim \nu} [E_i] \ge 2/3$ for some $i \in \{1, 2, 3\}$. If $i = 1$, set $u_a = (a_1, \vec{0}, -1/2) \in \mathbb{R}^{2d+1}$ where $\vec{0} = \underbrace{(0, \dots, 0)}_{d \text{ times}}$. If $i = 2$, set $u_a = (\vec{0}, a_2, -1/2)$. Finally, if $i = 3$, set $u_a = (\vec{0}, \vec{0}, 1)$. In any case, we have $\Pr_{b \sim \nu}[\sgn\langle u_a, v_b\rangle \ne A_{(a_1, a_2), (b_1, b_2)}] \le 1/3$ as desired.

    The sign rank lower bound for this matrix follows from~\Cref{thm:HS-lower-bound}.
\end{proof}

\bibliography{bibo}

\appendix

\section{Deferred Proofs} \label{app:deferred}

\subsection{Tail Bounds for Inner Products of Random Unit Vectors}
\label{app:tail-bounds}

\getkeytheorem{tail-bounds}

\begin{proof}
By rotational symmetry, $S$ has the same distribution as the first
coordinate of a uniform point on $\mathbb{S}^{d-1}$.
Let $G_1, \ldots, G_d$ be independent standard Gaussian variables, and put
$V = (G_2^2 + \cdots + G_d^2)^{1/2}$.
The direction of a Gaussian vector is uniform on the sphere, so $S$ is distributed identically to $\frac{G_1}{\sqrt{G_1^2+V^2}}$.

Using the density function of $G_1$, we get that for all $u < 1/2$,
\begin{equation} \label{eq:gauss-ex-identity}
  \E \left[ e^{uG_1^2} \right]
  = \frac{1}{\sqrt{2\pi}}\int_{-\infty}^\infty e^{ux^2} \cdot e^{-x^2/2}\, \dd x
  = (1-2u)^{-1/2}.
\end{equation}
Markov's inequality implies
\begin{align*}
    \Pr[|S|>\delta]
    &= \Pr\left[ (1-\delta^2)G_1^2>\delta^2V^2 \right] \\
    &= \Pr\left[ \exp \left( \frac{1-\delta^2}{2}G_1^2 - \frac{\delta^2}{2} V^2 \right) > 1 \right] \\
    &\leq \E \left[ \exp \left( \frac{1-\delta^2}{2}G_1^2 - \frac{\delta^2}{2} \sum_{j=2}^d G_j^2 \right) \right] \tag{Markov's inequality}.
\end{align*}
By independence of the $G_i$s,
\begin{align*}
    \E \left[ \exp \left( \frac{1-\delta^2}{2}G_1^2 - \frac{\delta^2}{2} \sum_{j=2}^d G_j^2 \right) \right] &= \E\left[ \exp\left( \frac{1-\delta^2}{2} G_1^2 \right) \cdot \prod_{j=2}^d \exp\left( -\frac{\delta^2}{2} G_j^2 \right) \right] \\
    &= \E\left[ \exp\left( \frac{1-\delta^2}{2} G_1^2 \right) \right] \cdot \prod_{j=2}^d \E\left[ \exp\left( -\frac{\delta^2}{2} G_j^2 \right) \right] \\
    &= \frac{1}{\delta} \cdot \left( 1 + \delta^2 \right)^{-(d-1)/2},
\end{align*}
where the last equality holds due to equation~\eqnref{eq:gauss-ex-identity}. Chaining the two inequalities together gives the upper tail bound of \eqnref{eq:score-upper-tail}.

For the lower bound on $\Pr[|S| \ge \delta]$, observe that $S$ is distributed identically to $\frac{G_1}{\sqrt{G_1^2 + V^2}}$. So the event $S \ge \delta$ is equivalent to $G_1 \ge \delta V / \sqrt{1-\delta^2}$.
Using the density function of the standard Gaussian, we have for every $t > 0$ that
\[\Pr[G_1 \ge t] = \int_0^\infty \frac{1}{\sqrt{2\pi}} e^{-(x+t)^2/2} \, \dd x \ge \frac{e^{-t^2}}{\sqrt{2\pi}} \int_0^\infty e^{-x^2} \, \dd x = \frac{e^{-t^2}}{2\sqrt{2}}.\]
Here, the inequality follows from the fact that $\frac{1}{2}(x+t)^2 \le x^2 + t^2$. Conditioning on $V$ and then taking expectations yields
\begin{align*}
    \Pr[S \geq \delta] &\ge \E_{V}\left[\Pr[G_1 \ge \frac{\delta V}{ \sqrt{1-\delta^2}} \mid V]\right] \\
    &\ge \frac{1}{2\sqrt{2}} \E_{V}[e^{-\delta^2V^2/(1-\delta^2)}] \\
    &\ge \frac{1}{2\sqrt{2}} \left(\frac{1-\delta^2}{1+\delta^2}\right)^{(d-1)/2} \\
    &\ge \frac{1}{4} (1-\delta^2)^{d-1}.
\end{align*}

The lower tail bound \eqnref{eq:score-lower-tail} is trivial when
$\delta \geq 1/2$.  Suppose $0 < \delta < 1/2$. 
Since $S$ is distributed identically to $\frac{G_1}{\sqrt{G_1^2 + V^2}}$, we have
$\Pr[|S| \leq \delta] = \Pr[|G_1| \leq \frac{\delta V}{\sqrt{1-\delta^2}}]$.
Since the probability density function of the standard Gaussian is at most $1/\sqrt{2\pi}$ everywhere, we have
\begin{align*}
  \Pr\left[ |G_1| \leq \frac{\delta v}{\sqrt{1 - \delta^2}} \mid V = v \right]
  \leq \frac{1}{\sqrt{2\pi}} \cdot \frac{2\delta v}{\sqrt{1-\delta^2}}.
\end{align*}
Note that
$\mathbb{E}[V] \leq (\mathbb{E}[V^2])^{1/2} = \sqrt{d-1}$.
So averaging over $V$ yields \eqnref{eq:score-lower-tail}:
\[
    \Pr[|S| \leq \delta] \leq \frac{\sqrt{2/\pi} \cdot \delta \cdot \sqrt{d-1}}{\sqrt{3/4}} \leq 2\delta\sqrt{d}.
\]
\end{proof}

\subsection{Properties of Gegenbauer Polynomials}
\label{app:gegenbauer}

\getkeytheorem{geg-facts}

\begin{proof} ~
    \begin{enumerate}
        \item Since the weight function $v_d(t) = (1-t^2)^{(d-3)/2}$ is even, we have
        \[\int_{-1}^1 t^i p(t) v_d(t) \, \dd t = 0\]
        whenever $i$ is even and $p$ is an odd function, and whenever $i$ is odd and $p$ is an even function. The claim then holds by induction, since the component of $t^i$ that remains after orthogonalizing it is odd if $i$ is odd and even if $i$ is even.

        \item To prove this identity, we will first show that $P_x(t)$ is a polynomial of degree $i$ in the variable $t$. We will then use its orthogonality to all lower degree polynomials to conclude that it must be a (rescaled) Gegenbauer polynomial.
        
        Fix a polynomial $q : \mathbb{S}^{d-1} \to \mathbb{R}$ as described and a point $x \in \mathbb{S}^{d-1}$. Below, fix an arbitrary extension of $q$ to a polynomial of degree at most $i$ in the ambient domain $\mathbb{R}^d$. Define a function $B$ over two scalar variables as
        \[B(r, s) = \mathop{\mathbb{E}}_{U \in \mathbb{S}^{d-1} : \langle U, x \rangle = 0}\left[q(rx + sU)\right].\]
        For every fixed $u$, the expression $q(rx + su)$ inside the expectation is a polynomial of degree at most $i$ in the variables $r, s$, so the function $B$ is as well. Moreover, $U$ has the same distribution as $-U$, so $B(r, s) = B(r, -s)$.
        This implies that $B$ only contains even powers of $s$. Hence, $P_x(t) = B(t, \sqrt{1-t^2})$ is a polynomial of degree $i$ in $t$.

        Now let $R(t)$ be an arbitrary polynomial on $[-1, 1]$ of degree at most $(i-1)$. We will show that $P_x$ is orthogonal to $R$ with respect to the weight $v_d$. To see this, we observe that
        \[\int_{-1}^1 P_x(t) R(t) v_d(t) \, \dd t = \left(\int_{-1}^1 v_d(t) \, \dd t\right) \cdot \mathop{\mathbb{E}}_{Y \in \mathbb{S}^{d-1}}\left[q(Y)R(\langle x, Y\rangle)\right] = 0.\]
        The last equality holds because $R(\langle x, y\rangle)$ is a polynomial in $y$ of degree at most $(i-1)$, and by our assumption on $q$. Hence, $P_x(t) = c \cdot G_{i}^{d}(t)$ for some constant $c$. This constant must be equal to $q(x)$ because $P_x(1) = q(x)$ and we are assuming the normalization $G_{i}^{d}(1) = 1$.

        \item Let $q$ be any nonzero polynomial satisfying the conditions of the spherical averaging identity of Part~2, e.g., one of the basis elements constructed in Section~\ref{sec:series-sym}. Let $M = \sup_{x \in \mathbb{S}^{d-1}} |q(x)| > 0$ and let $x$ be the point where this maximum is attained (such a point exists by continuity). Since $P_x(t)$ is an average of values of $q$ on the unit sphere, $|P_x(t)| \le M$ for every $t \in [-1, 1]$. Meanwhile, by Part~2, we have that $|G_{i}^{d}(t)| = |P_x(t) / q(x)| = |P_x(t)|/M$ for every $t \in [-1, 1]$, so this quantity always has absolute value at most $1$.

        \item Let $P = G_i^{d}$ and, for each $0 \le j \le i$, let $R_j = G_j^{d-2}$. Write $P = \sum_{j=0}^i \alpha_j R_j$ as a linear combination of the orthogonal basis elements $R_0, \dots, R_i$. Since $P(1) = R_0(1) = \dots = R_i(1) = 1$, we have that $\sum_{j = 0}^i \alpha_j = 1$. So to show this is a convex combination, it suffices to show that all of these coefficients are nonnegative.
        
        First observe by (anti-)symmetry (Part~1) that only those indices $j$ with parity equal to that of $i$ appear with nonzero coefficients. Thus, for all distinct remaining indices $j \ne j'$, the polynomial $R_j - R_{j'} = 0$ at $-1$ and $1$. This means we may write $R_j(t) - R_{j'}(t) = (1-t^2)\widetilde{R}(t)$ for some polynomial $\widetilde{R}$ of degree at most $i-2$. Since $(1-t^2)v_{d-2}(t) = v_d(t)$, we have
        \[\int_{-1}^1 P(t) (R_j(t) - R_{j'}(t)) v_{d-2}(t) \, \dd t = \int_{-1}^1 P(t) \widetilde{R}(t) v_{d}(t) \, \dd t = 0\]
        by orthogonality. Hence, the inner products $\int_{-1}^1 P(t) R_j(t) v_{d-2}(t)\ dt$ share some common value $c$ for all relevant indices $j$. The nonzero coefficients $\alpha_j$ in the orthogonal expansion of $P$ are given by
        \[\alpha_j = \frac{\int_{-1}^1 P(t) R_j(t) v_{d-2}(t) \, \dd t}{\int_{-1}^1 R_j^2(t) v_{d-2}(t)\, \dd t} = \frac{c}{\int_{-1}^1 R_j^2(t) v_{d-2}(t)\, \dd t}.\]
        Since each denominator is positive, $c$ must be positive, and hence all of the coefficients $\alpha_j$ are nonnegative.
    
        \item  Let $P = G_i^{d}$ and let 
        \[Q(t) = t P(t) - \frac{1-t^2}{d-3} P'(t).\]
        This is a polynomial of degree $i + 1$ with $Q(1) = 1$. Our goal is to show that $Q = G_{i+1}^{d-2}$, which must be the case if we can show that it is orthogonal to every polynomial of degree at most $i$ under the weight $v_{d-2}$. To see this, first observe by direct calculation that
        \[Q(t) v_{d-2}(t) = -\frac{1}{d-3} \frac{\dd}{\dd t} (P(t) v_d(t)).\]
        Let $R(t)$ be any polynomial of degree at most $i$. Integrating by parts yields
        \[\int_{-1}^1 Q(t) R(t) v_{d-2}(t) \, \dd t = \frac{1}{d-3} \left(P(-1)R(-1)v_d(-1) - P(1)R(1)v_d(1) + \int_{-1}^1 P(t) R'(t) v_d(t) \, \dd t\right).\]
        The boundary terms are $0$ because $v_d(1) = v_d(-1) = 0$ for $d \ge 4$. The integral is also $0$ by orthogonality, since $R'(t)$ has degree at most $i - 1$. Hence $Q$ is orthogonal to $R$ under the weight $v_{d-2}$.
    \end{enumerate}
\end{proof}

\subsection{Rational Approximations to the Sign Function}
\label{app:newman}

\getkeytheorem{newman}

\begin{proof}
    We follow Newman's argument closely \cite{newman1964approximation}.
    Suppose $p$ and $q$ are polynomials of degree at most $D$ such that $q$ is non-zero on $I_\eta$ and for all $x \in I_\eta$, $|\frac{p(x)}{q(x)} - \sgn x| \leq 1/k$.
    Let $A = p - q$ and $B = p + q$. Neither $A$ nor $B$ is identically zero, since otherwise $p/q$ would be identically $1$ or $-1$ and could not approximate the sign function.

    For every $x \in [\eta,1]$, the assumption gives
    \begin{align*}
         \frac{k-1}{k} &\leq \frac{p(x)}{q(x)} \leq \frac{k+1}{k}, \text{ and} \\
        -\frac{(k+1)}{k} &\leq \frac{p(-x)}{q(-x)} \leq -\frac{(k-1)}{k}.
    \end{align*}
    It follows that $B(x)$ and $A(-x)$ are nonzero on $[\eta, 1]$ and
    \begin{align*}
        \left| \frac{A(x)}{B(x)} \right| &= \frac{|1-p(x)/q(x)|}{|1+p(x)/q(x)|} \leq \frac{1}{2k-1}, \text{ and} \\
        \left|\frac{B(-x)}{A(-x)}\right| &= \frac{|1+p(-x)/q(-x)|}{|1-p(-x)/q(-x)|} \leq \frac{1}{2k-1}.
    \end{align*}
    Multiplying these inequalities, taking logarithms, and
    integrating gives
    \begin{equation} \label{eq:int-ab-upper}
        \int_\eta^1 \left( \ln \left| \frac{A(x)B(-x)}{B(x)A(-x)} \right| \right) \frac{\dd x}{x}
        \leq - 2\ln(2k-1) \ln\frac{1}{\eta}.
    \end{equation}
    While $A(x)B(-x)$ might have roots, the integral nonetheless converges when evaluated as an improper integral.

    Let $x$ be an arbitrary positive real number, and fix $z \in \mathbb{C}\setminus\{0\}$.
    By the triangle and reverse triangle inequalities, whenever $x \neq |z|$,
    \[
        \left|
            \ln\left|\frac{x-z}{x+z}\right|
        \right|
        \leq
        \ln\frac{x+|z|}{|x-|z||}.
    \]
    Substituting $t = x/|z|$ and integrating gives
    \begin{align*}
        \int_\eta^1 \left| \ln \left| \frac{x-z}{x+z} \right| \right|\frac{\dd x}{x}
        &\leq \int_\eta^1 \left( \ln \frac{x + |z|}{|x - |z||} \right) \frac{\dd x}{x} \\
        &\leq \int_0^\infty \left( \ln\frac{1+t}{|1-t|} \right) \frac{\dd t}{t}
        =\frac{\pi^2}{2}.
    \end{align*}
    Thus for all $z \in \mathbb{C} \setminus \{0\}$,
    \begin{equation}
        \int_\eta^1 \left( \ln \left| \frac{x-z}{x+z} \right| \right) \frac{\dd x}{x} \geq -\frac{\pi^2}{2}. \label{eq:log-z-int-ub}
    \end{equation}
    For $z=0$, the logarithm is identically zero, so the same bound holds.

    Let $z_1,\ldots,z_m$ and $w_1,\ldots,w_n$ be the complex roots of $A$ and $B$, respectively, counted with multiplicity.
    Since both polynomials have degree at most $D$, we have $m, n \leq D$. Factoring the polynomials and rearranging, we get
    \[
        \ln\left| \frac{A(x)B(-x)}{B(x)A(-x)} \right|
        = \sum_{i=1}^{m} \ln\left| \frac{x-z_i}{x+z_i} \right| - \sum_{i=1}^{n} \ln\left| \frac{x-w_i}{x+w_i} \right|.
    \]
    Integrating both sides and using inequality \eqnref{eq:log-z-int-ub},
    \[
        \int_\eta^1 \left( \ln \left| \frac{A(x)B(-x)}{B(x)A(-x)} \right| \right) \frac{\dd x}{x} \geq -\frac{(m+n)\pi^2}{2}\geq -D\pi^2.
    \]
    Combining this with the upper bound~\eqnref{eq:int-ab-upper} we get
    \[
        -D\pi^2 \leq -2\ln(2k-1) \ln\frac{1}{\eta},
    \]
    which yields the desired inequality.
\end{proof}

\subsection{Properties of (Shifted) Legendre Polynomials}
\label{app:legendre}

\getkeytheorem{legendre-props}

\begin{proof}
    Note that for all nonnegative integers $a, b$, repeated integration by parts gives
    \[
        \int_0^1 t^a (1-t)^b \,\dd t = \frac{b!}{(a+1)(a+2) \cdots (a+b)} \int_0^1 t^{a+b} \,\dd t = \frac{a!b!}{(a+b+1)!}.
    \]
    So by a change of variables ($t \gets t + 1$) we get that for all $m, j$,
    \begin{align*}
        \int_1^2 (t-1)^m L_j(t) \,\dd t &= \int_0^1 t^m L_j(t+1) \,\dd t \\
        &= \sum_{r=0}^j (-1)^{j-r} \binom{j}{r}^2 \int_0^1 t^{m+r} (1-t)^{j-r} \,\dd t \\
        &= \sum_{r=0}^j (-1)^{j-r} \binom{j}{r}^2 \frac{(m+r)! (j-r)!}{(m+j+1)!} \\
        &= \frac{m!j!}{(m+j+1)!} \sum_{r=0}^j (-1)^{j-r} \binom{j}{r} \binom{m+r}{m}.
    \end{align*}
    Now notice that by the binomial theorem, $\binom{m+r}{m}$ is the coefficient of $x^m$ in the polynomial $(1+x)^{m+r}$, and so the sum in the expression above is precisely the coefficient of $x^m$ in the polynomial
    \[
        P(x) = \sum_{r=0}^j (-1)^{j-r} \binom{j}{r} (1+x)^{m+r} = (1+x)^m (-1 + (1+x))^j = (1+x)^m x^j.
    \]
    In particular, this means that the sum equals 0 when $m < j$ and 1 when $m = j$.

    Let $i, j$ be arbitrary and assume without loss of generality that $i \leq j$.
    Since $L_i$ has degree $i$, there exist coefficients $a_0, \ldots, a_i$ such that $L_i(t) = \sum_{m = 0}^i a_m (t-1)^m$.
    If $i < j$, then by linearity
    \begin{align*}
        \int_1^2 L_i(t) L_j(t) \,\dd t &= \sum_{m=0}^i a_m \int_1^2 (t-1)^m L_j(t) \,\dd t \\
        &= \sum_{m=0}^i \left[ \frac{a_m \cdot m!j!}{(m+j+1)!} \sum_{r=0}^j (-1)^{j-r} \binom{j}{r} \binom{m+r}{m} \right] \numberthis \label{eq:inner-product-sums} \\
        &= 0,
    \end{align*}
    where the last equality holds since $m \leq i < j$ in all the inner sums.
    Note that this argument applies not just to $L_i$ but to all polynomials of degree less than $j$; this will be useful in proving item~2.

    Now suppose $i = j$.
    Then all inner sums except the one where $m = i = j$ are 0, and the inner sum with $m = i = j$ equals 1.
    Moreover, it follows from the definition of $L_i$ that $a_i = \sum_{r=0}^i \binom{i}{r}^2 = \binom{2i}{i}$.
    \begin{align*}
        \int_1^2 L_i(t) L_j(t) \,\dd t = \frac{a_i \cdot (i!)^2}{(2i+1)!} = \binom{2i}{i} \frac{(i!)^2}{(2i+1)!} = \frac{1}{2i+1}.
    \end{align*}
    Item~1 is hence proved.
    
    Item 2 is immediate when $i = 0$. Fix $i \in \N$.
    Set $w(t) = (t-1)(2-t)$ and $\lambda = i(i+1)$.
    We first show that $(wL_i')' + \lambda L_i \equiv 0$.
    Let $q = (wL_i')' + \lambda L_i$.
    It is easy to see that since the leading term of $w(t)$ is $-t^2$, the degree-$i$ terms in the definition of $q$ cancel and hence $q$ is a polynomial of degree strictly less than $i$.
    Integrating by parts twice and using the fact (implicit in the proof of item~1) that $L_i$ is orthogonal to all polynomials of degree less than $i$ on $[1,2]$,
    \begin{align*}
        \int_1^2 q(t)^2 \,\dd t &= \int_1^2 q(t) \cdot ((w(t)L_i'(t))' + \lambda L_i(t)) \,\dd t \\
        &= \int_1^2 L_i(t) \cdot ((w(t)q'(t))' + \lambda q(t)) \,\dd t = 0.
    \end{align*}
    The boundary terms while integrating by parts vanish since $w(1) = w(2) = 0$. Since $q(t)^2$ is a continuous and non-negative function, this means $q(t) = 0$.

    Now define $E(t) = \lambda L_i(t)^2 + w(t) L_i'(t)^2$.
    Note that on $[1,2]$, $\lambda L_i(t)^2 \leq E(t)$ since $w$ and $(L_i')^2$ are non-negative on this interval.
    We have
    \begin{align*}
        E' &= 2\lambda L_i L_i' + w' (L_i')^2 + 2wL_i' L_i'' \\
        &= 2L_i' \cdot (\lambda L_i + wL_i'' + w'L_i') - w' (L_i')^2 \\
        &= 2L_i' \cdot q - w' (L_i')^2 \\
        &= - w' (L_i')^2 = (2t-3) L_i'(t)^2.
    \end{align*}
    It follows that $E$ decreases on $[1,3/2]$ and increases on $[3/2,2]$, meaning that it attains its maximum on $[1,2]$ at one of the endpoints.
    It is easy to see from the definition of $L_i$ that $L_i(1) = (-1)^i$ and $L_i(2) = 1$, and so $E(1) = E(2) = \lambda$.
    Thus on $[1,2]$, $\lambda L_i(t)^2 \leq \max\{E(1), E(2)\} = \lambda$.
    Item~2 follows.

    We now turn to item~3.
    Let $s_1(m)$ be the sum of the magnitudes of the coefficients of $(t-1)^m$.
    By the binomial theorem, we have
    \begin{align*}
        (t-1)^m &= \sum_{\ell=0}^m \binom{m}{\ell} t^\ell (-1)^{m-\ell} \\
        \implies s_1(m) &= \sum_{\ell=0}^m \binom{m}{\ell} = 2^m.
    \end{align*}
    Similarly, let $s_2(m)$ be the sum of the magnitudes of the coefficients of $(t-2)^m$.
    We have
    \[
        s_2(m) = \sum_{\ell=0}^m \binom{m}{\ell} 2^{m-\ell} = 3^m.
    \]
    Let $S_i$ be the sum of the magnitudes of the coefficients of $L_i$.
    Using the elementary fact that the sum of magnitudes of coefficients of the product of two polynomials is at most the product of the sums of magnitudes of the individual polynomials, we get that
    \begin{align*}
        S_i &\leq \sum_{r = 0}^i \binom{i}{r}^2 s_1(r) s_2(i-r) \\
        &\leq 3^i \sum_{r = 0}^i \binom{i}{r}^2 \\
        &= 3^i \binom{2i}{i} \leq 3^i \cdot 4^i = 12^i.
    \end{align*}
\end{proof}

\end{document}